\documentclass[manuscript]{acmart}
\setcopyright{none}
\usepackage{textcomp}
\usepackage{xcolor}

\usepackage{amssymb}
\usepackage{times,enumerate,amsmath,subfigure,graphicx,color,ifthen,amsfonts,url}
\usepackage{algorithm,algpseudocode,stmaryrd,newfile,mathtools,bm}
\usepackage{mathtools}
\usepackage[parfill]{parskip}

\usepackage{listings}
\usepackage{tikz}

\definecolor{mygreen}{rgb}{0,0.2,0}
\definecolor{mygray}{rgb}{0.5,0.5,0.5}
\definecolor{mymauve}{rgb}{0.58,0,0.82}
\definecolor{mypurple}{rgb}{0.38,0,0.32}
\definecolor{myblue}{rgb}{0.1,0,0.32}
\newcommand{\costyle}{\footnotesize\ttfamily\bfseries}
\newcommand{\kwstyle}{\costyle\textcolor{myblue}}

\newcommand{\B}[1]{\bm{#1}}
\newcommand{\CC}[1]{\bm{\mathcal{#1}}}

\renewcommand{\Vec}[1]{\ensuremath{\bm{#1}}}                   % vector
\newcommand{\Mat}[1]{\ensuremath{\bm{#1}}}                     % matrix
\newcommand{\fnrm}[1]{{\| #1 \|}_F}

\title[Distributed-Memory Tensor Completion for Generalized Loss Functions using New Sparse Tensor Kernels]{Distributed-Memory Tensor Completion for Generalized Loss Functions in Python using New Sparse Tensor Kernels}
\author{Navjot Singh}
\email{navjot2@illinois.edu}
\author{Zecheng Zhang}
\author{Xiaoxiao Wu}
\author{Naijing Zhang}
\author{Siyuan Zhang}
\author{Edgar Solomonik}
\email{solomon2@illinois.edu}
\affiliation{%
  \institution{University of Illinois at Urbana-Champaign}
  \country{USA}
}

\begin{document}
\begin{abstract}
    Tensor computations are increasingly prevalent numerical techniques in data science, but pose unique challenges for high-performance implementation.
    We provide novel algorithms and systems infrastructure which enable efficient parallel implementation of algorithms for tensor completion with generalized loss functions.
    Specifically, we consider alternating minimization, coordinate minimization, and a quasi-Newton (generalized Gauss-Newton) method. 
    By extending the Cyclops library, we implement all of these methods in high-level Python syntax.
    To make possible tensor completion for very sparse tensors, we introduce new multi-tensor primitives, for which we provide specialized parallel implementations.
    We compare these routines to pairwise contraction of sparse tensors by reduction to hypersparse matrix formats, and find that the multi-tensor routines are more efficient in theoretical cost and execution time in experiments.
    %that can be computed with asymptotically less cost than by pairwise tensor contraction for key components of the tensor completion methods.
    %In particular, we show how TTTP can be used to perform a quasi-Newton step via conjugate gradient with implicit matrix-vector products in parallel.
    %Further, we also provide another multi-tensor routine, Solve Factor, which is useful for performing alternating minimization efficiently.
    %Using high-level Python syntax, we implement these algorithms, which fully automates the management of distributed-memory parallelism and sparsity for NumPy-style operations on multidimensional arrays.
    We provide microbenchmarking results on the Stampede2 supercomputer to demonstrate the efficiency of the new primitives and Cyclops functionality.
    We then study the performance of the tensor completion methods for a synthetic tensor with 10 billion nonzeros and the Netflix dataset, considering both least squares and Poisson loss functions.
\end{abstract}

%\keywords{tensor completion, alternating least squares, coordinate descent, stochastic gradient descent, Gauss-Newton method, Cyclops Tensor Framework.  }
\maketitle
%%% FOR ARXIV
\thispagestyle{empty}
%%% END FOR ARXIV

\section{Introduction}
\label{sec:intro}
Emerging sparse tensor methods pose new challenges for high-performance programming languages and libraries.
This paper describes new steps in making high-level productive parallel programming for sparse tensor algebra possible.
We focus specifically on formulation and implementation of optimization algorithms for tensor completion, which require management of extremely sparse tensors and complicated tensor operations. We provide algorithms and software for tensor completion with generalized loss functions \cite{hong2020generalized}. 
Algorithms for the least-squares loss tensor completion have been a target of recent parallel implementation efforts~\cite{KARLSSON2016222,Smith:2016:EOA:3014904.3014946}.
%We innovate in the formulation of the most prominent and complex tensor completion methods to simplify the required kernels.
We provide new formulations of a variety of tensor completion algorithms for the generalized loss function based on a common set of basic kernels. We implement these kernels to provide a new programming abstraction and software infrastructure for distributed-memory sparse tensor optimization algorithms. By extending Cyclops~\cite{solomonik2013cyclops}, we provide a Python-level interface to these sparse tensor kernels that achieve scalability on high performance distributed-memory architectures.

Tensor completion~\cite{liu2012tensor}, a generalization of the matrix completion problem, is the task of building a model to approximate a tensor based on a subset of observed entries.
The model should accurately represent observed entries, generalize effectively to unobserved entries, have a concise representation, provide efficient prediction of any tensor entry, and be possible to optimize.
Low-rank matrix factorizations are a widely used model for matrix completion, while tensor decompositions~\cite{doi:10.1137/07070111X}, especially the canonical polyadic (CP) decomposition~\cite{hitchcock1927expression,doi:10.1137/07070111X}, are commonly used for tensor completion~\cite{gemulla2011large}.
The major computational challenge in tensor completion is the optimization of the model, i.e., computation of a low-rank CP decomposition that effectively approximates the observed entries~\cite{KARLSSON2016222}.

We consider four optimization methods for tensor completion, described in detail in Section~\ref{sec:cmpl}.
Alternating minimization or alternating least squares for least-squares loss (ALS) updates one factor matrix while keeping other factor matrices fixed for each step, yielding a  symmetric positive definite (for convex loss functions) linear system of equations to be solved.
Coordinate minimization (coordinate descent) updates one column of a factor matrix while keeping others fixed for each step and alternates among factor matrices in a cyclic manner. 
Compared to ALS, coordinate descent performs updates with less computational cost, but reduces the minimization objective more slowly in each sweep of update.
Stochastic gradient descent (SGD) randomly selects samples from the tensor at each iteration and optimizes all the factor matrices based on these entries with a gradient-based update.

Second-order algorithms like Newton's method and Gauss-Newton method for CP decomposition with least-squares loss have been shown to perform better than ALS when the factors are highly correlated and an accurate solution is required \cite{sorber2013optimization,acar2011scalable}. However, each iteration of these algorithms is expensive and naive approaches do not scale due to the size of linear system required to solve. 
For the decomposition problem, the Gauss-Newton method can be implemented efficiently using an implicit form of the Hessian for fast matrix-vector products in the Conjugate gradient algorithm  \cite{sorber2013optimization,singh2019comparison}. Recently, a second order (Gauss-Newton like) method was proposed for the generalized decomposition problem~\cite{9145601}. In~\cite{9145601}, the structure of the Hessian of the generalized decomposition problem is explored and the method is shown to perform better than the gradient-based LBFGS~\cite{hong2020generalized} for beta divergence loss functions. In Section~\ref{sec:cmpl:GN}, we use tensor algebra to introduce a new formulation of the Newton's method and consequently a quasi-Newton method for generalized tensor completion which leverages an implicit form of the Hessian arising in the completion problem. Note that if the number of missing entries is set to zero, the problem would become a decomposition problem and the implicit form of the Hessian would correspond to the Hessian constructed in~\cite{9145601}. The implicit form of the Hessian can be leveraged to solve the linear system involving the large Hessian by use of batched conjugate gradient.  
The implicit matrix-vector products are easier to perform efficiently with basic sparse tensor algebra operations and the overall method achieves a lower computational cost than a direct solve~\cite{9145601,PAATERO1997223}.

To achieve high-performance for sparse tensor completion, we extend the functionality for sparse tensor contractions included in Cyclops~\cite{solomonik2015sparse,Solomonik:2017:SBC:3126908.3126971,solomonik2014massively}.
Since tensor completion is often done with extremely sparse tensor datasets~\cite{frosttdataset,liu2012tensor}, the use of CSR sparse matrix format for contraction of local tensor blocks becomes inefficient, and hypersparse matrix formats~\cite{buluc2008representation} are necessary.
We add support for a doubly compressed CSR format to Cyclops (described in Section~\ref{sec:krnl:hsparse}), which requires $O(m)$ memory for a tensor with $m$ nonzeros, and provides functionality to support contraction of a sparse and dense tensors (into a sparse output) using the hypersparse format.
Support of this format in a distributed memory library imposes new challenges, such as the necessity to perform summation and distributed reduction of blocks in hypersparse format.
To the best of our knowledge, Cyclops is the first distributed tensor library to offer this functionality.

We also identify a common generic multi-tensor routine that arises in tensor completion, and is likely to be very useful in generalized CP decomposition of sparse tensors as well as other applications.
This routine cannot be executed efficiently by the standard approach of pairwise contraction of tensors which is typically used by Cyclops.
Therefore, in Section~\ref{sec:krnl:tttp}, we introduce a programming abstraction for this tensor-times-tensor-product (TTTP) routine that achieves lower cost and memory footprint via a specialized parallel implementation.
Specifically, the TTTP routine multiplies entries of a sparse tensor with corresponding multilinear inner products of vectors.
Alternating minimization requires computation of tensor contractions with a sparse tensor along with solving systems on the fly to avoid overheads in memory footprint. To address this, we provide a programming abstraction for a sub-iteration of the algorithm involving specialised sparse tensor contractions. We develop library routines that map sparse tensor contractions to sparse matrix products and specialised multilinear operations.

We develop parallel implementations of the tensor completion methods leveraging a new Python interface to Cyclops (described in Section~\ref{sec:prog}).
This interface provides routines for Einstein-summation-like contraction of tensors, TTTP, and a multitude of other operations manipulating sparse and dense tensors.
The functionality is interfaced via Cython~\cite{behnel2010cython} to the C++ core of Cyclops.
Cyclops itself uses MPI, OpenMP, and CUDA to perform tensor algebra and data transformations/redistribution.
A basic set of parallel dense linear algebra routines are made available by interfacing to ScaLAPACK~\cite{Dongarra:1997:SUG:265932}.
The Python interface implements much of the basic functionality of \lstinline[language=Python]{numpy.ndarray}, making it possible to easily transform sequential Python dense tensor codes to distributed-memory-parallel sparse tensor software.

We provide performance results on the Stampede2 supercomputer for redistribution, tensor contractions, TTTP, and tensor completion algorithms with Cyclops.
Our results demonstrate that the new hypersparse representations enable contraction of tensors with extremely low density and that our new specialized TTTP algorithm achieves much better scalability than when done by pairwise contraction.
Finally, our tensor completion results show the capability of a high-level Python implementation of tensor completion methods to scale to tens of thousands of cores and 10B nonzeros of a highly sparse ($10^{-5}$ density) tensor. As an example for our algorithmic and software framework for generalized tensor completion, we provide an implementation of tensor completion algorithms for least-squares loss and the first distributed memory implementation of tensor completion algorithms for Poisson loss with logarithm link function~\cite{hong2020generalized}, and these methods are able to show good performance on the Netflix dataset~\cite{bennett2007netflix} for tensor completion.

This paper makes the following contributions:
\begin{itemize}
\item novel algorithms for alternating minimization and coordinate minimization for generalized tensor completion,
\item novel formulation of the second order algorithm for generalized tensor completion that uses implicit conjugate gradient and is easily implementable with tensor algebra kernels,
\item novel infrastructure for hypersparse matrix formats for general sparse--dense parallel tensor contractions,
\item a new programming abstraction for products of sparse tensors and tensor products (TTTP) and solving the linear systems arising in alternating minimization for generalized tensor completion (Solve Factor) ,%a new programming abstraction for solving linear systems arising in alternating minimization for generalized tensor completion
\item novel support of distributed-memory sparse tensor algebra operations in Python by a new interface to Cyclops,
\item first parallel implementation of generalized tensor completion algorithms that use second order information.
\end{itemize}

\section{Tensor Completion}
\label{sec:cmpl}
%\begin{itemize}
%\item Introduction to tensors and tensor algebra operations (contraction, tensor product, summation, broadcast, reduce, semiring operations)
%\item Tensor in CP format and Tensor decompositions \\
A tensor $\CC{T} \in \mathbb{R}^{I_1 \times \dots \times I_N}$ has \textbf{order} $N$ (i.e. $N$ modes/indices), \textbf{dimensions} $I_1 \text{-by-} \dots \text{-by-} I_N$ and \textbf{elements} $t_{i_1 \dots i_N} = t_i $ where $i \in \bigotimes^{N}_{i=1}\{1, \dots , I_i\}$.
Order $N$ tensors can be represented by $N-$dimensional arrays.
The algorithms and techniques involved in tensor completion do not differ significantly for tensors of order 3 or larger, and many tensor datasets are order 3, so we focus on this case for simplicity of presentation.

\subsection{Tensor Completion by generalized CP Decomposition}

The canonical polyadic (CP) decomposition~\cite{hitchcock1927expression} of an order three tensor $\CC{T} \in \mathbb{R}^{I \times J \times K}$ has the form,
\begin{equation}\label{eq:1}
t_{ijk} = \sum_{r=1}^R u_{ir}v_{jr}w_{kr},
%\boldsymbol{T}= \sum_{l=1}^{L} x^{(1)}_{l} \circ x^{(2)}_{l} \circ \dots x^{(N)}_{l}, \qquad x^{(n)}_{l} \in \mathbb{R}^{I_n}
%\boldsymbol{T}= \sum_{l=1}^{L} x^{(1)}_{l} \circ x^{(2)}_{l} \circ \dots x^{(N)}_{l}, \qquad x^{(n)}_{l} \in \mathbb{R}^{I_n}
\end{equation}
where $R$ is referred to as the \textbf{rank} of the decomposition and $\B U$, $\B V$, $\B W$ as \textbf{factor matrices}.
Letting $\langle \cdot,\cdot,\cdot\rangle$ denote a trilinear inner product, we can rewrite the above in terms of the rows $\Vec{u}_i$, $\Vec{v}_j$, $\Vec{w}_k$ of the factor matrices,
\begin{equation}\label{eq:2}
t_{ijk} = \langle \Vec{u}_i, \Vec{v}_j, \Vec{w}_k\rangle.
\end{equation}
%The symbol $\circ$ is the vector outer product.
%The \textbf{rank} of the tensor is the smallest $L$ possible.
The set of observed entries of a tensor $\CC T$ may be represented by index set $\Omega\subset \{1,\ldots, I\}\times \{1,\ldots, J\}\times \{1,\ldots,K\}$, so that for all $(i,j,k)\in\Omega$, $t_{ijk}$ has been observed.
%The task is to fill elements $x_l^{(n)}$ in \ref{eq:1} of a rank $L$ tensor.
%Let $P_{\Omega}$ be the projection of the known indices on the index set,
%$$\C V = P_\Omega(\C W) \Rightarrow v_{ijk}=\begin{cases} v_{ijk} = w_{ijk} & : (i,j,k) \in \Omega \\ v_{ijk} = 0 & : \text{otherwise} \end{cases}.$
The objective function that we seek to minimize is
%\begin{align*}
%\min_{\B U,\B V,\B W} 
%\underbrace{\sum_{(i,j,k)\in\Omega} \big(t_{ijk}{-}\langle \Vec{u}_i, \Vec{v}_j, \Vec{w}_k\rangle\big)^2}_{\text{Frobenius norm error on observed entries}} 
%\min \norm{P_{\Omega} \biggl(\Mat{T}-\sum_{l=1}^{L} x^{(1)}_{l} \circ x^{(2)}_{l} \circ \dots x^{(N)}_{l}\biggr)}_2^2 \\
%+ \underbrace{\lambda (\fnrm{\B U}^2 {+} \fnrm{\B V}^2 {+} \fnrm{\B W}^2).}_{\text{regularization to prevent overfitting}}
%\end{align*}
%In a more general setting, the objective function changes to $f(\B U,\B V, \B W)=$
\begin{equation}\label{eq:obj}
%\min_{\B U,\B V,\B W} 
f(\B U,\B V, \B W)=\underbrace{\sum_{(i,j,k)\in\Omega} \phi(t_{ijk},\langle \Vec{u}_i, \Vec{v}_j, \Vec{w}_k\rangle)}_{\text{Sum of elementwise loss function defined on observed entries}}{+} \underbrace{\lambda (\fnrm{\B U}^2 {+} \fnrm{\B V}^2 {+} \fnrm{\B W}^2).}_{\text{regularization to prevent overfitting}}
\end{equation}
where $\phi: \mathbb{R}\times \mathbb{R} \rightarrow \mathbb{R}$ is an arbitrary function which can be chosen according to the inherent data distribution. Various potential choices of these functions have been discussed in~\cite{hong2020generalized}.

Most of the previous work for CP tensor completion is related to least-squares loss function, i.e., by using the following elementwise function in the equation~\ref{eq:obj},
\[\phi(t_{ijk},\langle \Vec{u}_i, \Vec{v}_j, \Vec{w}_k\rangle) = (t_{ijk} -  \langle \Vec{u}_i, \Vec{v}_j, \Vec{w}_k\rangle)^2,\]
which assumes that the data is normally distributed.  However, many datasets that we encounter do not satisfy this assumption, but may fall in a different category, for example, a dataset of counts might follow Poisson distribution with the elementwise loss function,
\begin{align*}
\phi(t_{ijk},\langle \Vec{u}_i, \Vec{v}_j, \Vec{w}_k\rangle) =  \langle \Vec{u}_i, \Vec{v}_j, \Vec{w}_k\rangle - t_{ijk} \log\langle \Vec{u}_i, \Vec{v}_j, \Vec{w}_k\rangle, \quad \quad 
\text{where } \langle \Vec{u}_i, \Vec{v}_j, \Vec{w}_k\rangle > 0.
\end{align*}
For further subsections, we will use a shorthand notation for the function $\phi^{(t_{ijk})} : \mathbb{R} \rightarrow \mathbb{R}$ as $\phi_{ijk}$ which assumes that the first input to the corresponding binary input function, $\phi$, is $t_{ijk}$. And further, we  define 
\begin{equation}\label{eq:deriv_tnsr}
    \phi'_{ijk} = \frac{\partial \phi(t_{ijk},m_{ijk})}{\partial m_{ijk}}, \quad 
    \text{ where }  m_{ijk} = \langle \Vec{u}_i, \Vec{v}_j, \Vec{w}_k\rangle
\end{equation}   and $\phi''_{ijk}$ accordingly.

 We will be exploring tensor completion algorithms for the generalized loss functions and how these relate to the special case of least-squares loss. We introduce a table of derivative information in Table~\ref{tab:derivs} which will be used throughout the further subsections to formulate the algorithms. More detail on these expressions are provided in the Appendix~\ref{sec:additional}. Note that $\odot$ is the Hadamard/pointwise product. 
\begin{table}[ht!]
\centering
    
	\caption{First and second order derivative information}
	\begin{tabular}{ |l|c|c|}	
		\hline		
		\textbf{Derivatives} & \textbf{General loss function} &
		$\phi = \big(t_{ijk}{-}\langle \Vec{u}_i, \Vec{v}_j, \Vec{w}_k\rangle\big)^2$ \\ \hline	$\nabla f(\Vec{u_i})$ & $ \sum_{(j,k)\in \Omega_i} (\Vec{v}_j \odot \Vec{w}_k) \phi'_{ijk} + 2\lambda \Vec{u}_i $ & $2\sum_{(j,k)\in \Omega_i} (\Vec{v}_j \odot \Vec{w}_k) \Big( \langle \Vec{u}_i, \Vec{v}_j, \Vec{w}_k\rangle -t_{ijk} \Big) + 2\lambda \Vec{u}_i$ \\ \hline
		$\frac{\partial f(u_{ir})}{\partial u_{ir}}$  & $\sum_{(j,k)\in \Omega_i} v_{jr}w_{kr}\phi'_{ijk} + 2\lambda u_{ir}$ & $2 \sum_{(j,k)\in \Omega_i} v_{jr}w_{kr}\Big( \langle \Vec{u}_i, \Vec{v}_j, \Vec{w}_k\rangle -t_{ijk} \Big) + 2\lambda u_{ir}$ \\ \hline 
		$\Mat{H}_{f}(\Vec{u}_i)$  & $\sum_{(j,k)\in \Omega_i} (\Vec{v}_j \odot \Vec{w}_k)^T \phi''_{ijk} (\Vec{v}_j \odot \Vec{w}_k)
        + 2\lambda\Mat{I}$ & 2$\sum_{(j,k)\in \Omega_i}(\Vec{v}_j \odot \Vec{w}_k)^T  (\Vec{v}_j \odot \Vec{w}_k) + 2\lambda\Mat{I}$ \\ \hline
        $\frac{\partial^2 f(u_{ir})}{\partial u^{2}_{ir}} $ &
        $\sum_{(j,k)\in \Omega_i} v^2_{jr}w^2_{kr}\phi''_{ijk} + 2\lambda$ & $2\sum_{j,k}\hat{\Omega}_{ijk}v_{jr}^2w_{kr}^2 +2\lambda$\\
        \hline
        
	\end{tabular}
	\label{tab:derivs}
\end{table}
\subsection{Alternating Minimization (Alternating Least Squares)}
\label{sec:cmpl:ALS}
Alternating minimization works by fixing all except one factor matrix at a time and solving the optimization problem with respect to that factor matrix optimally. For solving the resulting subproblem, each row of the factor matrix can be optimized via Newton's method, 
\begin{align*}
\Vec{u}^{\text{(new)}}_i = \Vec{u}_i + \Delta \Vec{u}_i, \\
\text{where} \quad \Delta \Vec{u}_i\Mat{H}_{f}(\Vec{u}_i)&= -\nabla f(\Vec{u}_i).
\end{align*}
Alternating least squares (ALS) method is a standard algorithm for  CP decomposition with least-squares loss of tensors. Fixing all except one factor matrices results in a quadratic subproblem which can be solved with a single Newton's step. Defining $\hat{\Omega}_{ijk} = 1$ if $(i,j,k) \in \Omega$ and $0$ otherwise, and taking $t_{ijk}=0$ if $(i,j,k) \notin \Omega$, and using Table~\ref{tab:derivs} for gradient and Hessian values with an initial guess of zeros for each row of the factor matrix, i.e., $\Vec{u}_{i} =\Vec{0}$, the above equations can be expressed by sparse tensor contractions,
\begin{align*}
\sum_{r} u^{\text{(new)}}_{ir}(g^{(i)}_{rs}+\lambda \delta_{rs}) &= \sum_{j,k}v_{js} w_{ks} t_{ijk}, \\
\text{where} \quad
g^{(i)}_{rs}&= \sum_{j,k}v_{jr}w_{kr} \hat{\Omega}_{ijk} v_{js}w_{ks}.
\end{align*}
Given $m=|\Omega|$ observed values, solving the linear systems has cost $O(IR^3)$, forming the right-hand sides has cost $O(mR)$, and computing the matrices $\Mat{G}^{(i)}$ has cost $O(mR^2)$.
Contracting two tensors at a time to form each $\Mat{G}^{(i)}$ all at once incurs additional memory footprint, specifically,
\[O(\min(\text{median}(I,J,K)R^2,mR)+LR^2),\text{ where} \]
\begin{align*}
L=\text{median}(&|\{(j,k) : (i,j,k)\in\Omega\}|,|\{(i,k) : (i,j,k)\in\Omega\}|,\\
&|\{(i,j) : (i,j,k)\in\Omega\}|).
\end{align*}
For any general objective function, solving for one factor matrix results in an optimization problem which is dependent on the chosen loss function and might not be quadratic. Newton's method is provably convergent for convex functions but, may take more than one step unlike, least-squares loss. With the Hessian and gradient for general loss functions for each row $\Mat{u}_i$ defined as in Table~\ref{tab:derivs}, the same analysis as above can be applied for each Newton's step except for the fact that $\phi'_{ijk}$ and $\phi''_{ijk}$ tensors would be computed beforehand, which require $O(mR)$ computational cost and $O(m)$ memory.
\vspace{0.5cm}
For ALS, loss computation can be accelerated by using the left and right hand sides from the previous sub-iteration by using the following identity
\[\sum_{i,j,k \in \Omega} \Big(t_{ijk} - \sum_r u_{ir}v_{jr}w_{kr}\Big)^2 = \sum_{i,j,k \in \Omega} t^2_{ijk} + \sum_{i,j,k \in \Omega}\Big(\sum_{r}u_{ir}v_{jr}w_{kr}\Big)^2 -2\sum_{i,j,k \in \Omega}\sum_rt_{ijk}u_{ir}v_{jr}w_{kr}   \]
The first term can be easily computed in beginning of the algorithm. The last term can be computed using the right hand sides computed in the previous ALS sub-iteration by using
\[\sum_{i,j,k \in \Omega}\sum_rt_{ijk}u_{ir}v_{jr}w_{kr}= \sum_{i,r}u_{ir}\Big(\sum_{j,k \in \Omega_{i}}t_{ijk}v_{jr}w_{kr}\Big)=\sum_{i,r}u_{ir}p_{ir}. \]
where $\Mat{P}^{(i)}$ is the right hand side for the $i^{th}$ row of the previously solved factor matrix. The middle term can be computed from the previously computed left hand sides $\Mat{G}^{(i)}$ as follows,
\[\sum_{i,j,k \in \Omega}\Big(\sum_{r}u_{ir}v_{jr}w_{kr}\Big)^2 = \sum_{i,r,z}u_{ir}\Big(\sum_{j,k}v_{jr}w_{kr}\hat{\Omega}_{ijk}v_{jz}w_{kz}\Big)u_{iz} = \sum_i\sum_{r,z} u_{ir}g^{(i)}_{rz}u_{iz},\]
where $\Mat{G}^{(i)}$ is the left hand side for the $i^{th}$ row of the previously solved factor matrix. Therefore, after a sub-iteration, loss can be computed with $O(IR^2)$ cost.

\subsection{Coordinate Minimization (Coordinate Descent)}
\label{sec:cmpl:CCD}
Rather than updating the whole row of a factor matrix as in alternating minimization, coordinate minimization updates a single variable at a time while keeping the others fixed. A Newton's step for a single variable is given as
\begin{align*}
u_{ir}^{\text{(new)}}&= u_{ir} + \Delta u_{ir},\\
\text{where} \quad \Delta u_{ir}&= -\frac{\frac{\partial f}{\partial u_{ir}}}{\frac{\partial^2 f(u_{ir})}{\partial u^{2}_{ir}}}.
\end{align*}
For the least-squares loss, one Newton's iteration is sufficient, and an iteration of coordinate descent is analogous to a step of ALS with a rank $R=1$ CP decomposition.
The main advantage over general ALS is the lack of a need to solve systems of linear equations in coordinate descent as the step can be computed using the values from Table~\ref{tab:derivs},
\[\Delta u_{ir} =  \bigg(-\lambda u_{ir} + \sum_{(j,k)\in \Omega_i}v_{jr} w_{kr} \big(t_{ijk} - \langle \Vec{u}_i, \Vec{v}_j, \Vec{w}_k\rangle  \big)\bigg)/\bigg(\lambda+\sum_{(j,k)\in \Omega_i}v_{jr}^2w_{kr}^2\bigg).
\]
Similar to ALS, we can use an initial guess of zeros, i.e., $u_{ir} =0$ and define
\[\rho_{ijk}^{(r)} = \begin{cases}
    t_{ijk} - \langle \Vec{u}_i, \Vec{v}_j, \Vec{w}_k\rangle + u_{ir}v_{jr}w_{kr},   & \text{if }  (i,j,k) \in \Omega \\
    0 & \text{otherwise.}
\end{cases}\]
 The update can be expressed with sparse tensor contractions,
\[u_{ir}^\text{(new)} = \Delta u_{ir} =  \bigg(\sum_{j,k}v_{jr} w_{kr} \rho^{(r)}_{ijk}\bigg)/\bigg(\lambda+\sum_{j,k}\hat{\Omega}_{ijk}v_{jr}^2w_{kr}^2\bigg). 
\]
These contractions can be performed with $O(m)$ cost to update each $u_{ir}$ for all $i$, and $\rho^{(r+1)}_{ijk}$ can be obtained from $\rho^{(r)}_{ijk}$ with $O(m)$ cost.
Consequently, coordinate descent also requires $O(mR)$ cost to update all factor matrix entries, but has less parallelism and generally makes less progress than ALS since the updates to elements of factor matrix rows are decoupled.
Our CCD implementation alternates between factor matrices for each column update, which corresponds to the CCD++ ordering~\cite{yu2012scalable}.

For a general loss function, similar to alternating minimization, we can use the values of derivatives with respect to general elementwise functions for each element $u_{ir}$ from Table~\ref{tab:derivs}  and compute the Newton's step with the same computational cost as above by using the trilinear product $\langle \Vec{u}_i, \Vec{v}_j, \Vec{w}_k\rangle$ to compute $\phi'_{ijk}$ and $\phi''_{ijk}$ tensors with an additional cost of $O(m)$. Similar to alternating minimization, note that unlike the least-squares loss, it may take more than one Newton's step to converge for a general loss function.

\subsection{Stochastic Gradient Descent}

Instead of solving for a subset of variables at a time, one can solve for all the variables in an iteration using the first order or the gradient information. The simplest algorithm which uses gradient information for all the variables is gradient descent. The values for the derivative with respect to each element can be used from Table~\ref{tab:derivs} yielding the following update,
\[u_{ir}^{(\text{new})} = u_{ir} - \eta \frac{\partial{f}}{\partial u_{ir}}\]
with a cost of $O(mR)$.

Since more accurate updates with monotonic convergence guarantees can be obtained with similar cost via ALS or coordinate descent, gradient descent is generally less efficient for tensor completion.
However, stochastic gradient descent offers a framework in which the initial tensor can be sampled, leading to cost $O(SR+(I+J+K)R)$
(where $S$ is the sample size) for a sweep that updates all factor matrices.

Again, for a general objective function an extra computational cost of $O(m)$ ($O(S)$ in case of stochastic gradient descent) is required to compute $\phi'_{ijk}$ tensor to compute the gradient for each factor matrix. A distributed memory implementation of SGD for tensor decomposition with generalized loss functions was recently released~\cite{osti_1656940}. Apart from stochastic gradient descent, LBFGS is another gradient based method that has been explored for generalized tensor decomposition in~\cite{hong2020generalized}. We do not consider LBFGS in this work.

%Unlike ALS which minimizes the objective when updating each factor matrices, gradient descent only improves it. But due to the relatively lower cost of calculating the gradient, this method is still widely adopted. The gradient for updating $\Vec{u}_i$ is the partial derivative of $f(\Vec{u}_i)$ w.r.t. $\Vec{u}_i$: 
%\[\frac{\partial f(\Vec{u}_i)}{\partial \Vec{u}_i} = 2\big(\lambda \Vec{u}_i - \sum_{(j,k)\in \Omega_i} r_{ijk} \Vec{v}_j \Vec{w}_k \big)\]
%where $r_{ijk} = a_{ijk} - \Vec{u}_i  \Vec{v}_j \Vec{w}_k$ is the residual. The full update for $\Vec{u}_i$ would then be:
%\[\Vec{u}_i = \Vec{u}_i - \eta \frac{\partial f(\Vec{u}_i)}{\partial \Vec{u}_i} = \Vec{u}_i - 2 \eta \big(\lambda \Vec{u}_i - \sum_{(j,k)\in \Omega_i} r_{ijk} \Vec{v}_j \Vec{w}_k \big)\]
%Here $\eta$ is the step size. Instead of a full update, stochastic gradient descent(SGD) randomly selects $(i, j, k) \in \Omega$ and then updates $\Vec{u}_i$:
%\[\Vec{u}_i = \Vec{u}_i - 2 \eta \big(\lambda \Vec{u}_i / |\Omega_i| -  r_{ijk} \Vec{v}_j \Vec{w}_k \big)\]
%
%

\subsection{Second order algorithms (Newton and Gauss-Newton algorithms)}
\label{sec:cmpl:GN}
In the same regime of optimizing all variables at once, second order information can be used to obtain Newton's and Gauss-Newton algorithms for the generalized completion problem. To minimize the objective function in equation~\ref{eq:obj}, each iteration of the algorithm updates all the factor matrices by using the following update,
\begin{align*}
    [\Mat{U}^{(\text{new})},\Mat{V}^{(\text{new})},\Mat{W}^{(\text{new})}] = [\Mat{U},\Mat{V},\Mat{W}] + [\Delta\Mat{U},\Delta\Mat{V},\Delta\Mat{W}],\\
    \text{where} \quad
    [\Delta\Mat{U},\Delta\Mat{V},\Delta\Mat{W}] = -\Mat{H}_f^{-1}(\Mat{U},\Mat{V},\Mat{W}) \nabla f(\Mat{U},\Mat{V},\Mat{W}),
\end{align*}
where $\Mat{H}_f(\Mat{U},\Mat{V},\Mat{W})$ is the Hessian or the approximated Hessian and $\nabla f(\Mat{U},\Mat{V},\Mat{W})$ is the gradient for the objective function $f(\Mat{U},\Mat{V},\Mat{W})$
in equation ~\ref{eq:obj} with respect to all the factor matrices.

While the gradient can be computed efficiently, explicitly computing the Hessian or approximated Hessian ($\Mat{H}_f(\Mat{U},\Mat{V},\Mat{W})$) and storing it is extremely expensive as it is sparsity unaware requiring $O((I+J+K)^2R^2)$ memory. Moreover, directly inverting the matrix $\Mat{H}_f(\Mat{U},\Mat{V},\Mat{W})$ requires $O((I+J+K)^3R^3)$ which is practically infeasible for large scale tensors.

Alternatively, we explore the implicit form of the Hessian for the generalized completion problem to formulate Newton and quasi-Newton algorithm that use conjugate gradient (CG) with implicit matrix-vector products as applied in the CP decomposition with least-squares loss~\cite{singh2019comparison}. We assume Convex loss functions to ensure positive definite Hessian which can be relaxed by using a Krylov subspace method for other loss functions. The first row of $\Mat{H}_f(\Mat{U},\Mat{V},\Mat{W})$ can be divided into three blocks, 
\[h_{ilrs}^{(1,1)} = \sum_{(j,k) \in \Omega_i}v_{jr}w_{kr} \phi''_{ijk}\delta_{il}v_{js}w_{ks}  , \quad h_{ilrs}^{(1,2)}= \sum_{(j,k) \in \Omega_i} v_{jr}w_{kr}\phi''_{ijk}u_{is}\delta_{jl}w_{ks}+ \sum_{(j,k) \in \Omega_i} w_{kr}\phi'_{ijk}\delta_{rs}\delta_{jl}, \]
\[h_{ilrs}^{(1,3)}= \sum_{(j,k) \in \Omega_i}v_{jr}w_{kr}\phi''_{ijk}u_{is}v_{js}\delta_{kl} + \sum_{(j,k) \in \Omega_i} v_{jr}\phi'_{ijk}\delta_{rs}\delta_{kl}, \]
where $\delta_{ij}$ is the Kronecker-Delta function. Similarly, $\Mat{H}_f(\Mat{U},\Mat{V},\Mat{W})$ can be divided into three blocks for the second and third row which observe a similar structure. More detail on how these are derived can be found in the Appendix~\ref{sec:additional}. The Gauss-Newton method approximates the Hessian by excluding the additional term required in the off-diagonal blocks of the Hessian, resulting in a quasi-Newton method for generalized objective functions. With the implicit form of Hessian or approximated Hessian described above, the solve required for each iteration in the second order method can be accomplished by performing CG method with implicit matrix-vector products. Tensor contractions for updating the first factor matrix iterate in the implicit matrix-vector product inside CG iteration of each Newton's iteration are
\[\sum_{s,l}h_{ilrs}^{(1,1)}x^{(1)}_{ls} = \sum_{s}\sum_{(j,k) \in \Omega_i}v_{jr}w_{kr} \phi''_{ijk}v_{js}w_{ks}x^{(1)}_{is}, \]
\[\sum_{s,l}h_{ilrs}^{(1,2)}x^{(2)}_{ls} = \sum_{s}\sum_{(j,k) \in \Omega_i} v_{jr}w_{kr}\phi''_{ijk}u_{is}w_{ks}x^{(2)}_{js} + \sum_{(j,k) \in \Omega_i} w_{kr}\phi'_{ijk}x^{(2)}_{jr}, \]
\[\sum_{s,l}h_{ilrs}^{(1,3)}x^{(3)}_{ls} = \sum_{s}\sum_{(j,k) \in \Omega_i}v_{jr}w_{kr}\phi''_{ijk}u_{is}v_{js}x^{(3)}_{ks} + \sum_{(j,k) \in \Omega_i} v_{jr}\phi'_{ijk}x^{(3)}_{kr}, \]
\[\text{where, } \quad \sum_{n=1}^3\sum_{s,l}h_{ilrs}^{(1,n)}x^{(n)}_{ls}\] corresponds to the first block of the matrix vector product of the Hessian and factor matrices. The other two blocks of the matrix vector product can be computed similarly.

Each contraction of the type 
\[\sum_{s}\sum_{j,k}x_{jr}y_{kr}\hat{t}_{ijk}u_{is}v_{js}w_{ks},\] where $\hat{\mathcal{T}}$ is a sparse tensor, can be computed in $O(mR)$ cost by breaking it down into two contractions, 
\[ z_{ijk} = \hat{t}_{ijk}\sum_{s}u_{is}v_{js}w_{ks} \text{  and  }
a_{ir} = \sum_{j,k}z_{ijk}x_{jr}y_{kr}, \]
each of which costs $O(mR)$. Therefore, a CG step for solving a system in the quasi-Newton algorithm costs $O(mR)$. For Newton's algorithm, an additional contraction would be required for each off-diagonal block. This contraction costs $O(mR)$, and would require an additional memory overhead of $O(m)$ for storing $\phi'_{ijk}$.

The computation cost of the quasi-Newton algorithm is dominated by CG iterations. The number of CG iterations can be reduced by using the block diagonal part of the Hessian as a pre-conditioner~\cite{singh2019comparison}. However, storing the explicit inverse of the diagonal blocks of $\Mat{H}_f(\Mat{U},\Mat{V},\Mat{W})$ may still be a memory bottleneck for large tensors. Instead,the inverse of a diagonal block of $\Mat{H}_f(\Mat{U},\Mat{V},\Mat{W})$ can be applied with a cost identical to solving for the linear systems in a sub-iteration of alternating minimization algorithm introduced in Section~\ref{sec:cmpl:ALS}.

\section{New Sparse Tensor Kernels}
\label{sec:krnl}
The aforementioned tensor completion algorithms require sophisticated support for sparse tensor operations.
We extend the Cyclops library for tensor computations, which already includes support for sparse tensor contractions, reducing these to matrix multiplication with CSR format locally.
Cyclops leverages a cyclic data layout on multidimensional processor grids to achieve good performance and load balance for sparse tensors.
However, we observe two major bottlenecks within the sparse tensor algebra operations required in tensor completion that warrant extensions of functionality.

We describe new infrastructure for hypersparse matrix formats, leveraging a doubly-compressed format, which is a special case of the compressed sparse fiber (CSF) layout~\cite{smith2015tensor,smith2015splatt}.
%Hypersparsity support is useful in any tensor decomposition leveraging 
We apply this infrastructure to obtain TTM and MTTKRP implementations that require a minimal amount of memory and flops~\cite{kaya2015scalable,li2017model,hayashi2018shared,ballard2017communication}.
Further, we provide a specialized all-at-once implementation of MTTKRP that is competitive in performance with specialized MTTKRP libraries.
%While, specialized MTTKRP algorithms often perform the operation in an all-at-once manner, pairwise contraction is cost efficient when hypersparse matrix layouts are employed.
Additionally, we introduce a kernel for multiplication of a sparse tensor with multilinear inner products of vectors (TTTP), resulting in an output sparse tensor of the same size.
Our parallelization of the kernel leverages batching to achieve lower memory-footprint than previous work~\cite{Smith:2016:EOA:3014904.3014946,KARLSSON2016222}.
TTTP generalizes the sampled dense--dense matrix multiplication (SDDMM) kernel~\cite{canny2013big,nisa2018sampled,kjolstad2017tensor}, and is useful also for CP decomposition of sparse tensors. 

We also introduce a kernel for solving linear systems on the fly arising in alternating minimization of the generalized objective function for CP completion (given in equation~\ref{eq:obj})
involving sparse tensors. For the special case of least squares loss, we achieve a comparable performance to the state of the art library for performing ALS completion~\cite{Smith:2016:EOA:3014904.3014946}.

\subsection{Hypersparse Matrix Formats}
\label{sec:krnl:hsparse}

Tensor contractions can be reduced to matrix multiplication with matrices that have the same number of sparse entries.
However, while it is uncommon in sparse matrix computations for entire rows or columns of a sparse matrix to be zero, the sparse matrix--matrix products occurring by reduction from tensor computations often have this property~\cite{smith2015splatt}.
A canonical example is the product of a sparse tensor and a dense matrix, which can be used an initial step for MTTKRP, yielding an intermediate that can typically be reused in multiple MTTKRP operations via dimension trees~\cite{kaya2018parallel} (also see~\cite{phan2013fast,vannieuwenhoven2015computing,ballard2017communication}).
In this tensor times matrix (TTM) operation, given an order three tensor, we seek to compute
\[z_{ijr} = \sum_k t_{ijk} w_{kr},\]
where $\CC{T}$ is sparse and $\Mat{W}$ is dense.
By merging $i$ and $j$ into a single index, TTM reduces to a matrix-matrix product of sparse and a dense matrix.
For $\CC{T}\in\mathbb{R}^{I\times J\times K}$, if the number of entries in $\CC{T}$ is less than $IJ$, then the above matricization of $\CC{T}$ is necessarily hypersparse (contains rows with only zero entries), and $\CC{Z}$ is sparse.
For many sparse tensor datasets, one of the modes is small, or the number of nonzeros scales with mode size, i.e., $m=O(I+J+K)$.
In both cases, we may obtain a matricization that is very hypersparse (most rows are zero), in which case the matricization of $\CC{Z}$ cannot be stored in a dense format without increasing memory footprint.

Cyclops represents static sparse tensor data in a COO-like format, storing a single 64-bit integer for each value to encode its global location in the tensor, with index-value pairs sorted locally.
When a contraction is executed, the locally stored portion of the tensor is transformed into a sparse matrix format.
We extend this mechanism to support a `CCSR' layout, which is a speical case of DCSR~\cite{buluc2008representation} and CSF~\cite{smith2015splatt}, where CSR is used to encode the nonzero rows only and an additional array is stored that maps nonzero rows to the original set of rows.
This layout requires $\Theta(m)$ storage if a tensor has $m$ nonzeros, improving on $\Theta(IJ+m)$ needed for plain CSR for the TTM operation above.
Multiplication of a CCSR matrix by a dense matrix is easy, it suffices to multiply the reduced CSR matrix by the dense matrix, then generate a new CCSR matrix to represent the sparse output, resulting in $O(mR)$ cost.
%, if $\Mat{W}\in\mathbb{R}^{K\times R}$.

Realizing CCSR functionality for arbitrary tensor contractions also necessitates implementation of sparse format conversions, summation of CCSR blocks, and interprocessor reduction.
We provide kernels for each of these steps.
When sparsity is involved, Cyclops first ensures that each index arising in the tensor contraction expression occurs in exactly two tensors.
If an index occurs in only a single tensor, pre- or post- processing can be performed to reduce or map the input or output, respectively.
If an index occurs in all three tensors (specifying a set of independent contractions), Cyclops duplicates the index, converting one of the sparse operands to a tensor of one order higher, placing the original data on the diagonal (e.g. $c_i=v_iw_i$ with sparse $\Vec{v}$ is performed via $\Vec{c}=\hat{\Mat{V}}\Vec{w}$ where $\hat{v}_{ii}=v_i$).
By ensuring that each index occurs in exactly two tensors, Cyclops is able to map the local part of the contraction to a matrix--matrix product.
Cyclops puts local parts of the tensor into sparse matrix format by first converting to COO then
%the local nonzeros of a sparse tensor, where each nonzero is stored as a pair containing a global 64-bit integer index and a value, into a coordinate (COO) matrix format.
%The COO format is then converted 
to CCSR format (for a standard sparse format, conversion to CSR works similarly).

Local summation of CCSR matrices requires identifying which rows are nonzero in both matrices, which is done by comparing the two sets of nonzero row indices.
The summation of each row is done by leveraging a dense array.
In particular, if each local matrix has $K$ columns, nonzeros in that row are accumulated to the corresponding entries of an array of size $K$, then the sparse sum is read back and the entries used are zeroed out.
The cost of this operation for summing each row scales with the number of nonzeros in the output row, but the buffer must be allocated and cleared, creating a potential bottleneck if the local sparse matrices are very hypersparse in both rows and columns (most rows and most columns are entirely zero).
For sparse tensor times matrix contractions arising in the tensor completion kernels, each column contains nonzeros.

\begin{figure}[t]
{

\centering
\includegraphics[width=3.6in]{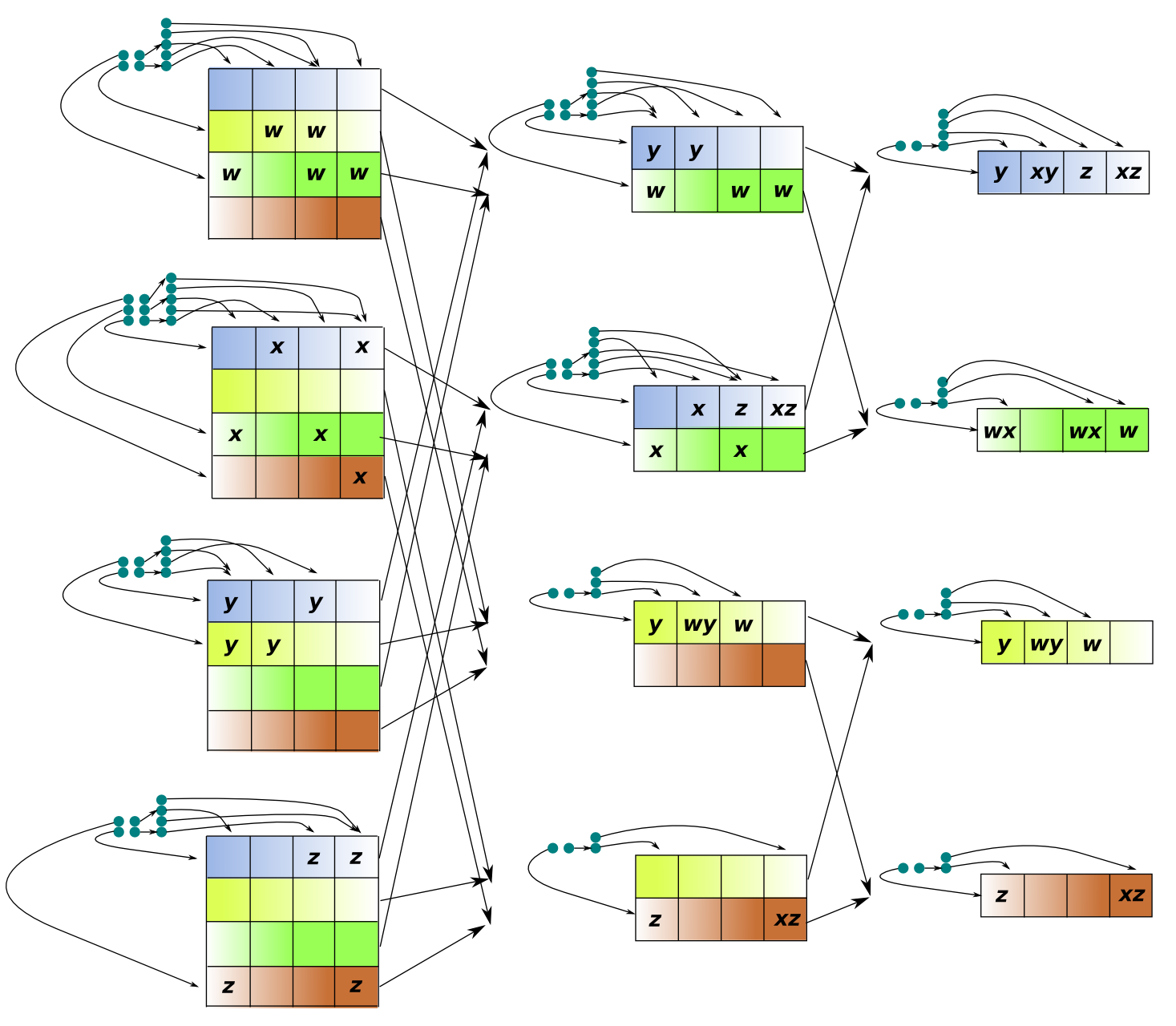}

}
\caption{Depiction of 4 processor reduce-scatter of $4\times 4$ hypersparse matrices stored in doubly compressed (CCSR) format.}
\label{fig:sparse_red}
\end{figure}

Parallel reduction of CCSR matrices leverages this summation kernel, using a butterfly collective communication approach (recursive halving followed by recursive doubling~\cite{thakur2005optimization}) that performs a sparse reduce-scatter followed by a sparse gather.
At each step of the sparse reduce-scatter, hypersparse matrices with smaller overall dimensions but higher density are summed by each processor using the sparse summation kernel described above.
An example of the reduce-scatter is displayed in Figure~\ref{fig:sparse_red}.
The sparse gather recombines these matrices by concatentation.
The partitioning and recombination is done using a $k$-ary butterfly, where $k$ is a parameter that we chose to be a constant.

\subsection{Matricized Tensor Times Khatri-Rao Product}
\label{sec:krnl:mttkrp}

While the use of hypersparse formats enables an implementation of MTTKRP that asymptotically minimizes memory footprint and cost, we also provide a specialized MTTKRP implementation that performs the operation in an all-at-once manner.
In particular, the MTTKRP is parallelized by performing smaller local MTTKRPs on each processor, using the sparse tensor data stored on that processor.
This parallelization follows SPLATT~\cite{smith2015tensor,smith2015splatt} and also uses a reduction to accumulate results.
However, the MTTKRP kernel interoperates with other Cyclops functionality, redistributing factor matrices from an arbitrary initial layout, to a partially-replicated distribution necessary to compute the local MTTKRP, and the resulting matrix is put into a layout that is distributed over all processors.
Locally, the sparse tensor data is kept in the usual Cyclops COO-like format, as opposed to the specialized CSF format.
Partial sums are accumulated for the local part of each tensor fiber along the most quickly changing index.
The BLAS \texttt{axpy} operation and the MKL vector pointwise product are used to achieve vectorization when the MTTKRP is performed with factor matrices that have more than one column.

\subsection{Tensor Times Tensor Product}
\label{sec:krnl:tttp}

Efficient support for sparse tensor contractions does not suffice for tensor completion algorithms.
Their use entails significant overhead in memory footprint even to just compute the residual,
\[\rho_{ijk} = t_{ijk} - \sum_{r=1}^R\hat{\Omega}_{ijk} u_{ir}v_{jr}w_{kr},\]
since forming intermediate or $x_{ijkr}=\hat{\Omega}_{ijk}u_{ir}$ increases memory footprint, while alternatively forming the dense intermediate $y_{ijk} = \sum_{r=1}^R u_{ir}v_{jr}w_{kr}$ is suboptimal in both memory footprint and work.
Evidently, the most efficient way to perform such operations requires all-at-once contraction of multiple operands.
To handle this operation effectively, we introduce the tensor-times-tensor product (TTTP) operation, which takes as input a sparse tensor $\CC{S}\in\mathbb{R}^{I_1\times \cdots \times I_N}$ and a list of up to $N$ matrices $\Mat{A}^{(1)}\in\mathbb{R}^{I_1\times R},\ldots,\Mat{A}^{(N)}\in\mathbb{R}^{I_N\times R}$ and computes
\[x_{i_1\ldots i_N} = s_{i_1\ldots i_N} \sum_{r=1}^R\prod_{j=1}^N a_{i_jr}^{(j)}.\]
If fewer then $N$ matrices are specified, the product should iterate only over modes for which an input is provided.
By iterating over $m$ nonzero entries in $\CC{S}$ and performing the multilinear inner product for each one, TTTP can be performed with cost $O(mR)$ and $O((I_1+\cdots +I_N)R+m)$ memory footprint.
When $N=2$, TTTP corresponds to the SDDMM operation $\Mat{X} = \Mat{S}\odot(\Mat{U}\Mat{V}^T)$.

TTTP is an integral part of the algorithms for  generalized tensor completion as it is used to compute $\phi'_{ijk}$ (equation~\eqref{eq:deriv_tnsr}) and $\phi''_{ijk}$. For the traditional tensor completion with least squares loss,
TTTP allows calculation of the residual in tensor completion with CP decomposition, by computing
\[\hat{\Omega}_{ijk}\sum_{r=1}^Ru_{ir}v_{jr}w_{kr}.\]
Although, this residual calculation can be accelerated in ALS as described in Section ~\ref{sec:cmpl:ALS}, it is explicitly necessary in the coordinate minimization algorithm.
Further, for the Newton's and quasi-Newton method with implicit conjugate gradient in Section~\ref{sec:cmpl:GN}, we use TTTP to compute updates via
\[z_{ijk} = \underbrace{\phi''_{ijk} \sum_{s} v_{js}w_{ks} x_{is}}_{\text{TTTP}},\quad x^{(new)}_{ir}= \underbrace{\sum_{j,k}v_{jr}w_{kr}z_{ijk}}_{\text{MTTKRP}}.\]
%TTTP also has use cases in performing CP decomposition (as opposed to completion) of a sparse tensor with $m$ nonzero entries $t_{ijk}$, $\forall (i,j,k)\in \Omega$.
%The routine provides an efficient way to compute the Frobenius norm of the residual~\cite{Smith:2016:EOA:3014904.3014946}, since
%\begin{align*}
%\sum_{i,j,k}&(t_{ijk}-\sum_{r=1}^R u_{ir}v_{jr}w_{kr})^2 \\
%=&\sum_{i,j,k}(t_{ijk}-\sum_{r=1}^R u_{ir}v_{jr}w_{kr})(t_{ijk}-\sum_{s=1}^R u_{is}v_{js}w_{ks})\\
%=&  \bigg(\sum_{r=1}^R\sum_{s=1}^R\bigg(\sum_{i} u_{ir}u_{is}\bigg)\bigg(\sum_{j} v_{jr}v_{js}\bigg)\bigg(\sum_{k} w_{kr}w_{ks}\bigg)\bigg)^2 \\
%{-}& \sum_{(i,j,k)\in\Omega}(\sum_{r=1}^R u_{ir}v_{jr}w_{kr})^2 
%{+} \sum_{(i,j,k)\in\Omega}(t_{ijk}{-}\sum_{r=1}^R u_{ir}v_{jr}w_{kr})^2.
%\end{align*}
%Specifically, TTTP efficiently computes $\sum_{r=1}^R\hat{\Omega}_{ijk} u_{ir}v_{jr}w_{kr}$, with which the residual terms in the final form can be obtained with cost $O(m+(I+J+K)R^2)$.
%These use cases generalize naturally to higher order tensors.
\begin{figure}[t]
{

\centering
\includegraphics[width=3.4in]{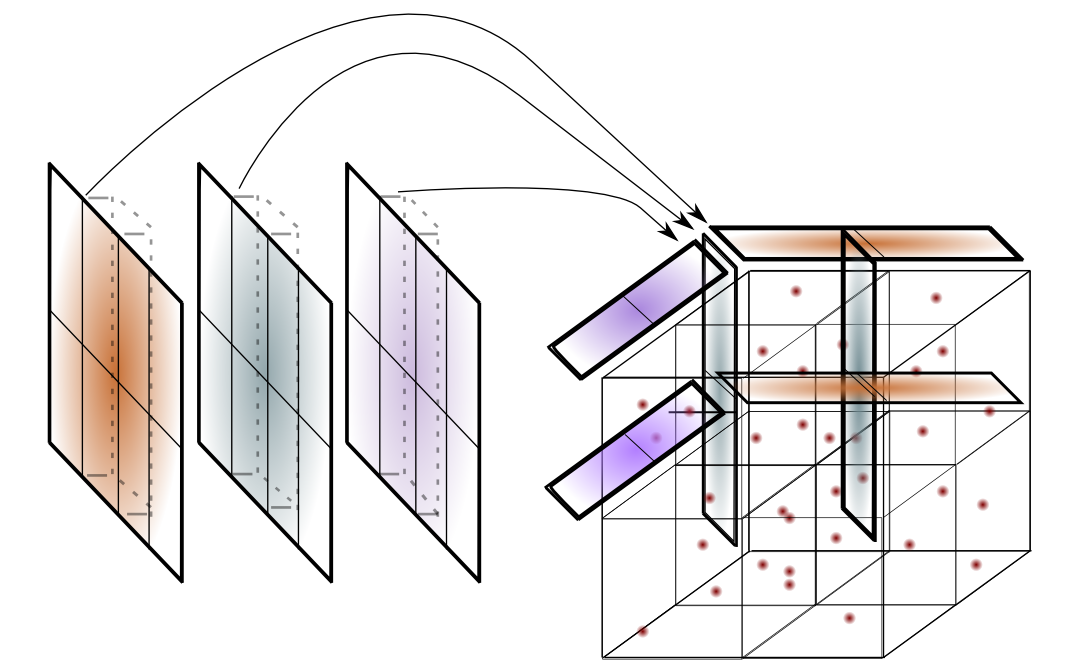}

}
\caption{Depiction of 8 processor parallelization of TTTP computing one of four smaller TTTP substeps.}
\label{fig:diag_tttp}
\end{figure}
Our parallel implementation of TTTP keeps the sparse tensor input $\CC{S}$ and output $\CC{X}$ local on whichever processor grid $\CC{S}$ was initially distributed on.
The matrices $\Mat{A}^{(1)},\ldots,\Mat{A}^{(N)}$ are input from an arbitrary initial processor grid distribution.
Each matrix is sliced into $H\leq R$ pieces by taking $H$ equal-sized subsets of their columns, based on available memory.
The computation then proceeds in $H$ steps, each computing a smaller TTTP involving matrices of size $I_j\times (R/H)$.
For each step, the corresponding slice of each of the $N$ matrices $\Mat{A}^{(j)}$ is redistributed so that its rows are cyclically distributed over the processor grid dimension along which the $j$th mode of $\CC{S}$ is distributed (if any), and replicated over all others.
Each of $P$ processors can then compute a part of the smaller TTTP with the entries of $\CC{S}$ (and $\CC{X}$) it is assigned locally, performing a total of $O(mR/P)$ work overall.

This parallel TTTP algorithm is depicted in Figure~\ref{fig:diag_tttp} for scenario with $P=8$ processors.
Assuming a $I=I_1=\cdots =I_N$ and a processor grid is used of dimensions $P^{1/N}\times \cdots\times P^{1/N}$, using a BSP model of communication~\cite{skillicorn1997questions,valiant1990bridging}, the latency cost (number of supersteps) is $O(H)$, the interprocessor bandwidth cost is $O(IR/P^{1/N})$, and the memory footprint is $O(m/P + IR/(P^{1/N}H))$.
Efficient mechanisms for redistribution of dense matrices between arbitrary processor grids exist in Cyclops~\cite{solomonik2014massively}.
%The columns of the matrices $\Mat{A}^{(1)},\ldots,\Mat{A}^{(N)}$ are each sliced into $H\leq R$ pieces, based on

\subsection{Solve Factor}
\label{sec:krnl:solvefac}

Alternating minimization for generalized CP tensor completion requires tensor contractions along with a solve which should be done on the fly to avoid a memory bottleneck. To accomplish this, we provide a specialized kernel, which uses a similar parallelization strategy suggested in~\cite{Smith:2016:EOA:3014904.3014946} for ALS completion. Our kernel takes as input a tensor $\CC{S}\in\mathbb{R}^{I_1\times \cdots \times I_N}$, a list of up to $N$ matrices $\Mat{A}^{(1)}\in\mathbb{R}^{I_1\times R},\ldots,\Mat{A}^{(N)}\in\mathbb{R}^{I_N\times R}$, an integer $n$, a right hand side matrix $\Mat{M}\in\mathbb{R}^{I_n\times R}$, and solves for the Newton's step with respect to $n^{\text{th}}$ factor matrix as described in Section~\ref{sec:cmpl:ALS}.

The left hand sides for $i_n^{th}$ row of the factor matrix in the Newton's step, $\Mat{G}^{(i_n)}$, can be computed by the following contractions, 
\[g^{(i_n)}_{rs} = \sum_{i_1 \dots i_{n-1}, i_{n+1} \dots i_N}\bigg(\prod_{p =1 , p \neq n}^Na_{i_pr}\bigg)s_{i_1 \dots i_N}\bigg(\prod_{p =1 , p \neq n}^Na_{i_ps}\bigg), \]
which together incur a computational cost of $O(mR^2)$ and a memory footprint of $O(I_nR^2)$.

Our parallel implementation follows the same strategy as in Section~\ref{sec:krnl:tttp} and keeps the sparse tensor input $\CC{S}$ local on whichever processor grid it was initially distributed on. The matrices $\Mat{A}^{(1)},\ldots,\Mat{A}^{(N)}$ are input from an arbitrary initial processor grid distribution and are redistributed as described in Section~\ref{sec:krnl:tttp}. To make use of BLAS-3 operations for the above mentioned contractions, the input tensor must be sorted with respect to the mode $n$. We accomplish this in the current format by performing a Counting sort~\cite{cormen2001introduction} using the indices of the $n^{\text{th}}$ mode as keys over the local data with a computational cost of $O(m/P)$.

Forming left hand sides for the normal equations for all the rows can be a memory bottleneck due to a memory footprint of $O(IR^2/P^{1/N})$. We divide the rows into $b$ batches according to the memory available, leading to a memory footprint of $O(IR^2/(bP^{1/N}))$ and then for computing the normal equations for each row, we store the hadamard products of the vectors multiplied with the square root of corresponding tensor entries in a local buffer of size $K \times R$ by using MKL for elementwise vector products. When the buffer is filled up or the number of entries are exhausted, a symmetric rank-k (SYRK) update is performed using BLAS to compute the local left hand
sides. A reduce scatter along slice with respect to mode $n$ of the processor grid allows us to scatter the computed left hand sides. The corresponding right hand sides are distributed and a symmetric positive definite solve routine in BLAS (POSV) is used to achieve a parallel solve with $O(IR^3/(bP^{\frac{N+1}{N}}))$ cost for a batch of rows.

%Note that in the current implementation, the elementwise function should be a convex function as we use a Positive Definite system routine to solve the linear system. This constraint is imposed as the Newton's method may fail to converge (or may take a long time to converge) for non-convex elementwise function and is a reasonable assumption as most of the loss functions used in practice are convex. However, this assumption can be easily relaxed by using a general linear system solve routine to achieve the local solve.  

\section{Python Interface and Implementation}
\label{sec:prog}
Cyclops~\cite{solomonik2014massively} provides extensive support for tensor algebra and tensor data manipulation in C++, leveraging BLAS~\cite{lawson1979basic}, MPI~\cite{Gropp:1994:UMP:207387}, OpenMP, CUDA, HPTT~\cite{springer2017hptt}, and ScaLAPACK~\cite{Dongarra:1997:SUG:265932}.
The library supports both dense tensor formats~\cite{solomonik2014massively} as well sparse tensor formats~\cite{solomonik2015sparse}, both of which leverage partitioning of the tensor data among all processors.
Scaling, summation, and contraction are supported via a succinct programmatic Einstein summation notation.
Cyclops also provides general kernels such as tensor transposition, redistribution, slicing, and permutation of tensor indices.
Additionally, the library supports user-defined element types and algebraic structures specifying their properties, as well as contractions that operate on tensors of different types, enabling applications such as graph algorithms~\cite{Solomonik:2017:SBC:3126908.3126971}.

Cyclops leverages a runtime-centric execution model, making data distribution and algorithmic scheduling decisions at execution time.
This enables performance models to be evaluated for runtime-determined parameters such as problem size and processor count.
We leverage this characteristic of the Cyclops system architecture to provide a performance-efficient Python interface to Cyclops.
This extension enables productivity for high-performance implementation of tensor computations.
By implementing a back-end for high-level NumPy-style operations\cite{van2011numpy}, we activate support for sparsity in tensor storage and computations, as well as parallel execution in distributed and shared memory.
These capabilities are enabled with minimal overhead to the user.
For example, by using Cyclops, sparse storage for a code based on the standard \lstinline[language=Python]{numpy.ndarray} can be implemented simply by an additional boolean flag in the \lstinline[language=Python]{ctf.tensor} constructor.
By contrast, the standard approach for supporting sparse matrix operations in Python, involves manual handling of the CSR format via SciPy~\cite{jones2014scipy}.

%As tensor algorithms become important in big data analysis and Python emerges as one of the most popular languages in the field of data analysis, we are motivated to provide a Python extension to CTF.

%\subsection{Cyclops Tensor Framework}
%
%\subsection{Python Extension Motivation}
%
%The tensor computation is prevalent in big data analysis, distributed-memory or the automated parallelization functionality of tensor computation becomes crucial to reduce the computational cost.
%With recent emerging of Python prevalence in data science, we are motivated to provide Python extension to the Cyclops tensor algebra library.
%As NumPy gives a standard multidimensional representation of arrays and interface of numerical operations \cite{van2011numpy} in Python, our extension to CTF has the numpy-style representation on tensors and tensor operations.
%There are some existing numpy-style packages which support the sparse multidimensional array computation and parallel complements to NumPy such as SciPy\cite{jones2014scipy}.

\subsection{Cyclops Python Interface}
\begin{figure}
    \centering
    \includegraphics[width=0.4\textwidth]{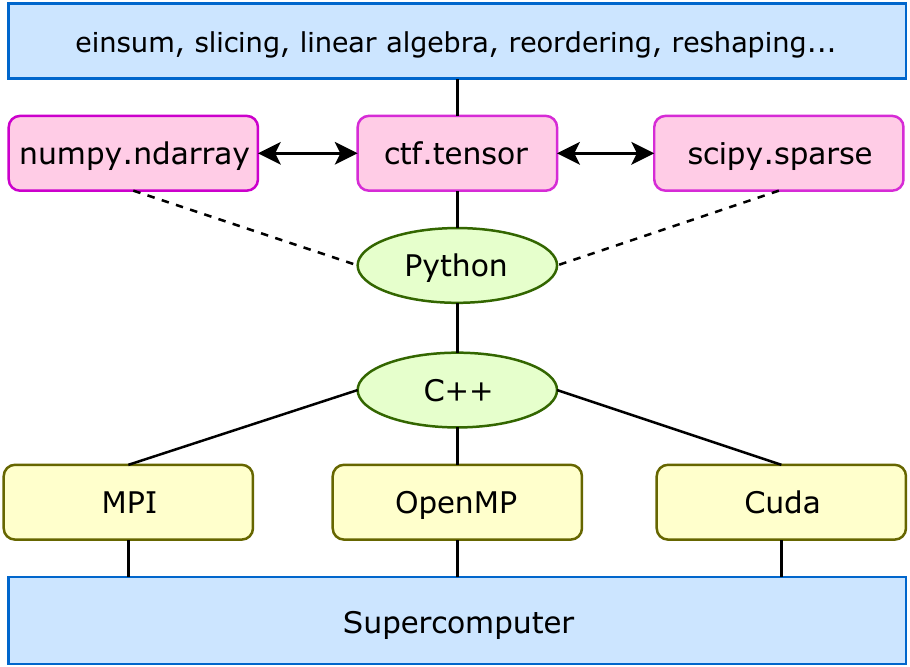}
    \caption{Overview of Cyclops Python interface organization.}
    \label{fig:ctf-architecture}
\end{figure}

We utilize Cython~\cite{behnel2010cython}, which enables interoperability of Python and C++, to encapsulate the main functionalities of Cyclops C++ interface.
As shown in Figure \ref{fig:ctf-architecture}, we introduce a Python tensor class that wraps the C++ Cyclops tensor object via Cython and provides the core functionality.
%The tensor class calls the tensor constructor in C++ interface once it is initialized.
Tensor and multidimensional array operations are also built on C++ interface functionalities including \lstinline[language=Python]{ctf.einsum}, \lstinline[language=Python]{ctf.tensordot}, \lstinline[language=Python]{ctf.transpose}, and  \lstinline[language=Python]{ctf.reshape}.
Functionality provided by NumPy in \lstinline[language=Python]{numpy.linalg} is also supported, including QR, Cholesky, SVD, and the symmetric eigensolve.
%Most of these functions support the sparse tensor calculation.
Boolean, integer, and floating point types of a variety of precision are supported, which are specified via \lstinline[language=Python]{numpy.ndarray.dtype}.
The C++ interface of Cyclops uses templating to support arbitrary types and user-defined elementwise operations, so extension of the Python interface to other types is possible.
Dense and sparse distributed Cyclops tensors may be defined in a variety of ways.
\begin{lstlisting}[language=Python, caption={Example Code: Tensor Initialization}, captionpos=b, label={lst:tensor}]
import ctf
U = ctf.tensor([5,7]) # dense zero matrix
M = ctf.random.random((4,4)) # random dense tensor
O = ctf.ones((4,3,5)) # tensor full of ones
I = ctf.eye(9) # dense identity
T = ctf.tensor([5,3,4], sp=True) # sparse tensor
T.fill_sp_random(-1.,1.,.1) # 10% density
S = ctf.speye(9) # sparse identity
\end{lstlisting}
%As shown in Example Code \ref{lst:tensor}, the tensor A is created with specified type  \lstinline[language=Python]{numpy.ndarray.float64} and dimensions three by four.
%Different from NumPy, sparse tensor can be specified directly by  the flag \lstinline[language=Python]{sp=True}.
For both the dense tensor and sparse tensor, NumPy-style indexing/slicing is provided such as \lstinline[language=Python]{A[0, 1]} to extract $a_{01}$ or \lstinline[language=Python]{A[3:5, 1:4:2]} to extract a $2$-by-$2$ matrix containing entries at the intersection of rows $3$ and $4$ and columns $1$ and $3$.
A key difference between the Cyclops Python interface and NumPy functions including slice and (transpose \lstinline[language=Python]{A.T}) is that Cyclops explicitly creates the new tensor in memory as opposed to providing a logical reference.
For example with the Cyclops interface, transposition is done via \lstinline[language=Python]{B = A.T()}, which returns a new tensor (so modifying elements of B will not change A).

Cyclops supports both NumPy-style Einstein summation, as well as an additional Einstein syntax similar to its C++ interface.
For example, the following two lines are equivalent.
\begin{lstlisting}[language=Python, caption={Example Code: Einstein Summation}, captionpos=b, label={lst:einsum}]
R += T - ctf.einsum("ir,jr,kr->ijk",U,V,W)
R.i("ijk") << T.i("ijk")- U.i("ir")*V.i("jr")*W.i("kr")
\end{lstlisting}
Expressions such as the above are passed directly to the C++ layer.
The C++ layer then makes decisions regarding evaluation ordering and choice of intermediate tensors.
Cyclops performs this by considering all possible binary trees for contraction of window of up to 8 tensors (contracting-away one tensor and including the next one given more than 8 operands), based on a heuristic model of computation and memory-bandwidth cost.
Intermediate tensors are defined to be sparse if they are a contraction of two sparse operands or if a very sparse tensor is contracted with a dense tensor (contraction corresponds to a matrix--matrix product with a hypersparse matrix that must have fewer than $1$ in $3$ rows with a nonzero).
%\begin{lstlisting}[language=Python, caption={Example code: NumPy Style Functions}, captionpos=b, label={lst:numpy}]
%    A = ctf.tensor([3,4], dtype=np.float64, sp=True)
%    A.T()
%    A.reshape([4, 3])
%    summation = A.sum()
%\end{lstlisting}

%Some NumPy functions are useful in scientific computing.
%In Example Code \ref{lst:numpy}, we show some NumPy style functions which are supported in the Python extensions.
%These functions support the sparse flag which can speed up the sparse tensor computation.

\subsection{TTTP Interface}

Cyclops does not automatically determine when to use the multi-tensor TTTP operation.
Instead, a simple interface is provided for this operation.
For example, the following code computes 
\begin{align*}
s_{ijkl}&=\sum_{r} o_{ijkl}u_{ir}v_{jr}w_{kr}z_{lr}, \quad
t_{ijkl}=\sum_{r} o_{ijkl}u_{ir}w_{kr}.
\end{align*}
\begin{lstlisting}[language=Python, caption={Example code: TTTP}, captionpos=b, label={lst:tttp}]
O = ctf.tensor((I,J,K,L),sp=True)
U = ctf.tensor((I,R)), V = ctf.tensor((J,R))
W = ctf.tensor((K,R)), Z = ctf.tensor((L,R))
... # fill O,U,V,W,Z
S = ctf.TTTP(Omega,[U,V,W,Z])
T = ctf.TTTP(Omega,[U,None,W,None])
\end{lstlisting}
The routine alternatively accepts a list of vectors rather than matrices as the second argument.
A similar routine is available via the C++ interface to Cyclops.

\subsection{Parallel Tensor Completion in Python}

Given high-level tensor algebra primitives, we are able to implement the aforementioned tensor completion algorithms without any explicit management of parallelism or data distribution.
The problem of parallelization of these algorithms is reduced to expressing them with high-level tensor algebra operations.

%We support core functionality for tensor completion, including the \lstinline[language=Python]{einsum} for tensor contraction and parallel MPI reader to read from the tensor file.
%The interface of \lstinline[language=Python]{einsum} is similar to that of NumPy.
%
%
%\begin{lstlisting}[language=Python, caption={Example code: Einsum}, captionpos=b, label={lst:einsum}]
%    ctf.einsum('ij->i', A)
%\end{lstlisting}
%
%As shown in Example Code \ref{lst:einsum}, the \lstinline[language=Python]{einsum} function takes the summation of each axes.
%
%Apart from the \lstinline[language=Python]{ctf.einsum}, a function called \lstinline[language=Python]{ctf.TTTP} takes a list of matrices or vectors and computes either in $B_{ijk} = A_{ijk}u_{i}v_{j}w_{k}$ or $B_{ijk} = \sum_{a}A_{ijk}u_{ia}v_{ja}w_{ka}$.
%Shown in Example Code \ref{lst:tttp}, the input is a list of matrices or vectors but can omit one of $u, v, w$ etc.
%We implement this function is that this kind of computation is core for the tensor completion task.
%
%\begin{lstlisting}[language=Python, caption={Example code: TTTP}, captionpos=b, label={lst:tttp}]
%    ctf.TTTP(A, [u, v, w])
%    ctf.TTTP(A, [u, None, w])
%\end{lstlisting}
%
%% \begin{itemize}
%% \item Summarize previous work on Cyclops
%% \item Motivate distributed-memory/parallel + sparse tensor functionality support for Python
%% \item Describe architecture/implementation of Cyclops Python extensions
%% \item Describe core functionality needed for tensor completion and optimization problems
%% \end{itemize}
\begin{figure*}[ht]
\centering
\begin{subfigure}[Tensor Transposition with Cyclops]
{\label{fig:transp}\includegraphics[width=0.45\textwidth]{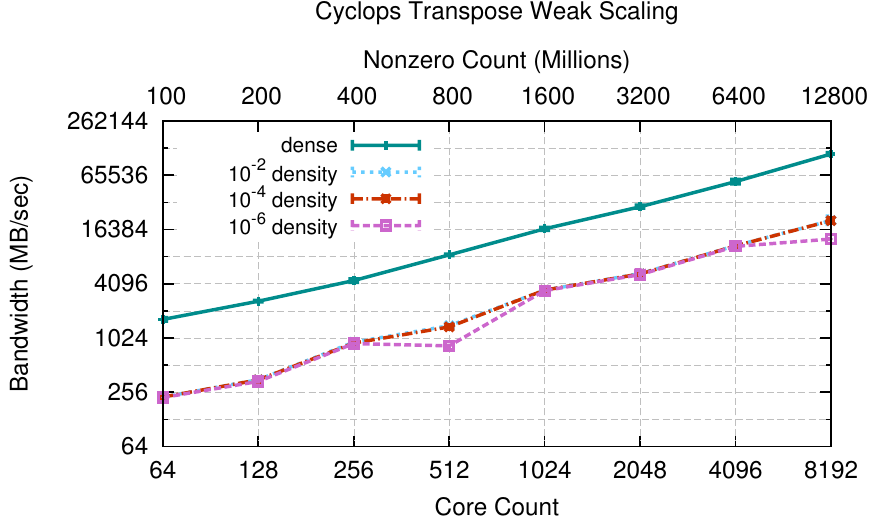}
}\end{subfigure}
\begin{subfigure}[Tensor Reshape with Cyclops]
{\label{fig:reshape}\includegraphics[width=0.45\textwidth]{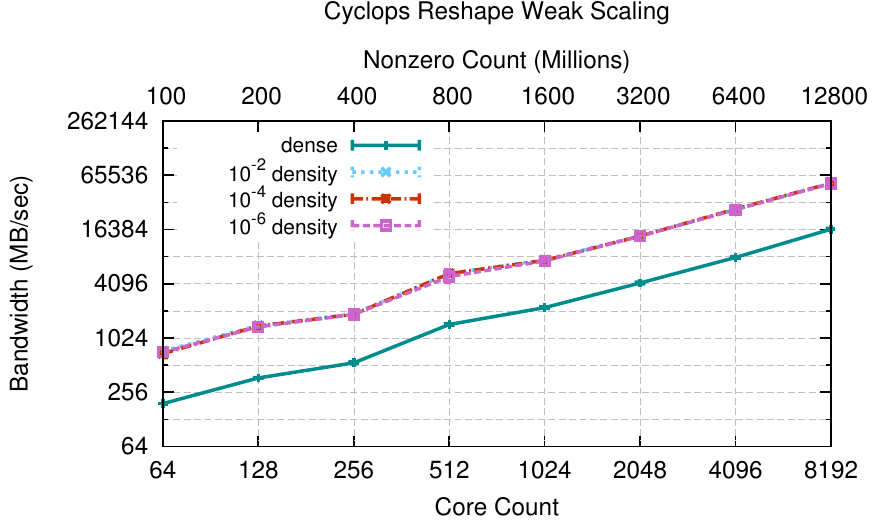}
}\end{subfigure}
\caption{Achieved bandwidth/throughout of transpose and reshape Python functions with Cyclops (16 bytes assumed for each nonzero in a sparse tensor and 8 bytes for each value in a dense tensor).}
\end{figure*}

\subsection{Alternating minimization (Alternating Least Squares) Implementation}
\label{sec:ALS_imp}
%In this study, previously proposed parallel algorithm that distribute the variables and let each processor independently update its own subset of the variables are implemented using CTF Python. 
%When updating rows of factor matrix, the most time-consuming tasks, unless the rank $R$ is very large, are the matrix product when forming $\Mat{G}$, and the matrix-vector product $\sum_{(j,k)\in \Omega_i}\Vec{v}_j \Vec{w}_k \Mat{\Omega}_{ijk} \Vec{v}_j \Vec{w}_k$. 
%To exploit the algebraic advantages of Python CTF in reducing memory cost and sparse format, we propose an implicit algorithm for parallel alternating least square. 
%In this algorithm, linear systems of equations are solved implicitly using implicit conjugate gradient rather than explicitly.
%The ALS algorithm can be implemented entirely using algebraic operations provided via the Cyclops interface.
Alternating minimization algorithm can be implemented entirely using the MTTKRP kernel for the right hand sides and the Solve Factor kernel for the solves.
%A key step within the algorithm are the implicit matrix--vector products within the batched conjugate gradient method~\eqref{eq:als_imp}.
%This step is implemented leveraging TTTP as well as Einstein notation for contraction in a single line of code.
%Specifically, the left hand side of the system of equations to be solve is formed using \lstinline[language=Python]{ctf.TTTP}:
Solving for one factor matrix can be implemented easily via the Cyclops Python interface.
%\begin{lstlisting}[language=Python, caption={ALS Implicit Batched CG Mat-Vec}, captionpos=b, label={lst:als}]
%Y.i("ir") << V.i("jr")*W.i("kr")* \
             %ctf.TTTP(Omega,[X,V,W]).i("ijk")
%\end{lstlisting}
\begin{lstlisting}[language=Python, caption={ALS solve for one factor matrix (\Mat{U})}, captionpos=b, label={lst:als}]
rhs = ctf.tensor((I,R))
ctf.MTTKRP(T,[rhs,V,W],0) #compute right hand sides
U = ctf.tensor((I,R))
ctf.Solve_Factor(Omega,[U,V,W],rhs,0,regu) #solve for U
\end{lstlisting} 
%The sparse tensor contractions are faster using \lstinline[language=Python]{ctf.TTTP} because it directly compute the multilinear products for each entry in the sparse tensor. 
%The error threshold of each implicit conjugate gradient iteration is a parameter that can be set to reduce computation time and improve performance.
%In our experiments, this parameter was statically set to $10^{-4}$.
%When updating rows, the implicit and explicit conjugate gradient are interchangeable.
%The implementation also leverages slicing to permit blocks of rows of factor matrices to be updated at a given time, but our experiments use only one block to maximize parallelism.
%are updated in blocks where there is a tradeoff between the number of blocks.  

\subsection{Coordinate Minimization (Coordinate Descent) Descent Implementation}

%The CCD++ formula of updating a column is described in the previous section, and here is the implentation in CTF Python on the update rule regarding column $f$ of $U$ (and similarly for $V$ and $W$):
The coordinate minimization updates are easy to formulate via Einstein notation contractions and elementwise operations.
\begin{lstlisting}[language=Python, caption={Example code: CCD++ Update Rule}, captionpos=b, label={lst:ccd}]
a = ctf.einsum('ijk,j,k->i',R,V[:,r],W[:,r])
b = ctf.einsum('ijk,j,j,k,k->i', Omega,V[:,r],V[:,r],W[:,r],W[:,r])
U[:,r] = a / (lmbda + b)
\end{lstlisting}
For the second expression above, Cyclops finds the right tree of contractions automatically (note that a tree is more efficient than contracting left-to-right given any initial order).
Slicing permits easy access of columns, although in our final implementation, we split up each factor matrix into column vectors outside of the CCD++ iteration loop to minimize overhead.

We also consider an implementation of CCD++ that is based on the MTTKRP kernel in Cyclops.
%TTTP routine in combination with a sparse summation operation.
This approach forgoes the need for tensor contractions with hypersparse matrix representations.
\begin{lstlisting}[language=Python, caption={Example code: CCD++ with MTTKRP}, captionpos=b, label={lst:ccd}]
ctf.MTTKRP(R,[A,V[:,r],W[:,r]],0)
ctf.MTTKRP(Omega,[B,V[:,r]*V[:,r],W[:,r]*W[:,r]], 0)
\end{lstlisting}
\subsection{Stochastic Gradient Descent Implementation}

%Commonly, SGD samples one entry at a time which impractical for the scale of the tensors used in this study. Instead we proposed a sampled subgradient descent method to achieve better efficiency in the CTF implementation. In each iteration, we sample index set $\Omega_{S} \in \Omega$ and apply updates to all the factor matrices alternatively. For updating $\Vec{u}$, we fix $\Vec{v}, \Vec{w}$ and apply:
%\[\Vec{u}_i = \Vec{u}_i - 2 \eta \big(\lambda \Vec{u}_i |\Omega_{S_i}|/|\Omega_{i}|- \sum_{(j,k)\in \Omega_{S_i}} r_{ijk} \Vec{v}_j \Vec{w}_k \big)\]
We leverage a sampling function in the Cyclops Python interface to obtain a random sample of the tensor $\CC{T}$ for each SGD sweep (update to each factor matrix).
%we can simply apply a dedicated sample function to any tensor $T$ to obtain a batched sample:
\begin{lstlisting}[language=Python, caption={Example code: SGD Batched Sampling}, captionpos=b, label={lst:sgd}]
sampled_T = T.copy()
sampled_T.sample(samprate)
sOmega = getOmega(sampled_T)
R = sampled_T - ctf.TTTP(sOmega,[U,V,W])
ctf.MTTKRP(R,[U,V,W],0)
U+= -2* step* lmbda *samprate*U
\end{lstlisting}
The bulk of the computation within SGD is then comprised of the above sparse MTTKRP, which calculates a subgradient from $\CC{R}$ (the residual for the sampled entries).
The \lstinline[language=Python]{getOmega()} function works by reading the local nonzeros of the tensor, and writing them to a new sparse tensor with unit values.
We also consider an implementation of SGD with the all-at-once Cyclops MTTKRP.

\subsubsection{Quasi-Newton (Gauss-Newton) Implementation} For implementing the quasi-Newton or Newton's algorithm, right hand sides in each iteration can be easily computed as these are negative of gradient with respect to each factor matrix. For solving the linear system in each iteration, CG iterations require matrix vector products with the implicit form of the Hessian. Each block of contraction required for the method as described in Section~\ref{sec:cmpl:GN} can be implemented using TTTP and the MTTKRP kernel. The output of TTTP is fed into the MTTKRP kernel with the desired output index. The expression for $(1,2)$ Hessian contractions is as follows 
\begin{lstlisting}[language=Python, caption={GN implicit block (1,2) contraction}, captionpos=b, label={lst:gn}]
A[0] += ctf.MTTKRP(ctf.TTTP(Omega,[U,Delta[1],W]), [None,V,W], 0)
\end{lstlisting}

Preconditioning can also be easily incorporated using the Solve Factor kernel used in Section~\ref{sec:ALS_imp}.

\section{Experimental Evaluation}
\label{sec:exp}

\begin{figure*}[ht]\centering
\begin{subfigure}[Tensor-Times-Matrix (TTM) with Cyclops]{\label{fig:ttm}\includegraphics[width=0.45\textwidth]{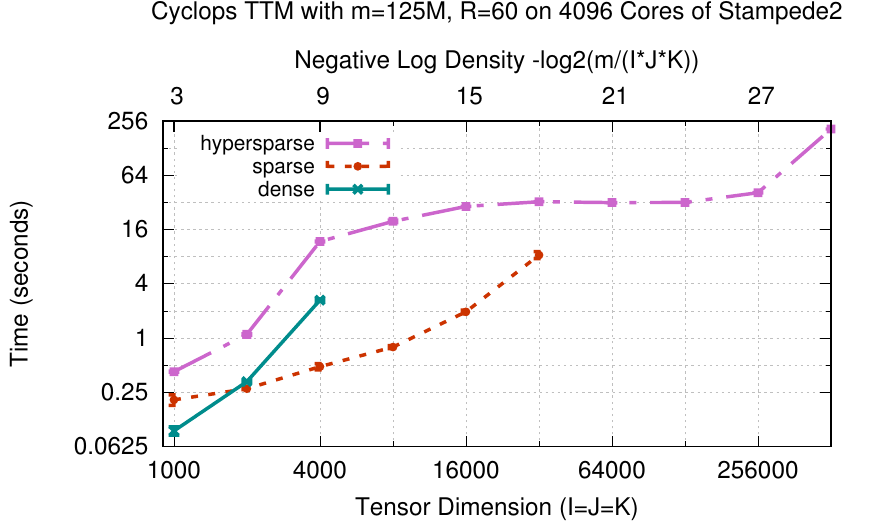}}
\end{subfigure}
\begin{subfigure}[MTTKRP with Cyclops]{\label{fig:mttkrp}\includegraphics[width=0.45\textwidth]{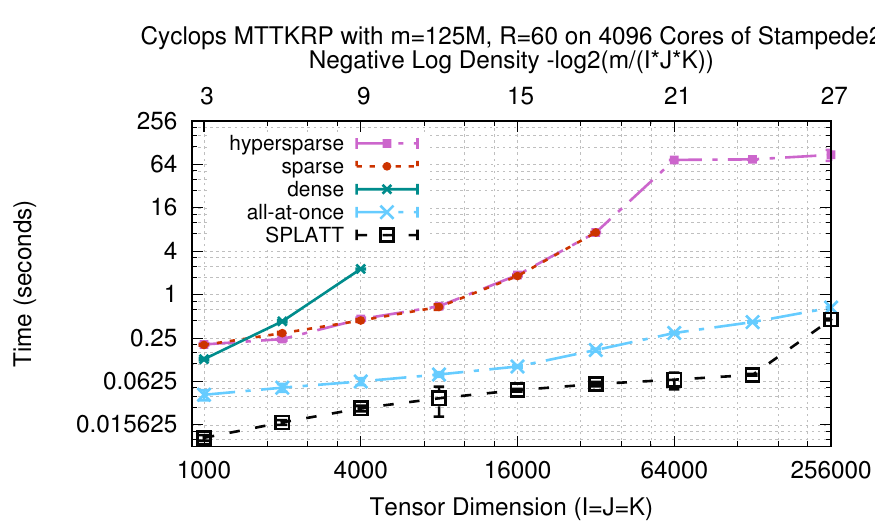}}
\end{subfigure}
\caption{Execution time of TTM and MTTKRP for order 3 tensors, both averaged over three possible variants (choices of contracted and uncontracted modes, respectively).}
\end{figure*}

\begin{figure*}[t]
{\centering
\begin{subfigure}[TTTP with $R=1$]
{\includegraphics[width=0.45\textwidth]{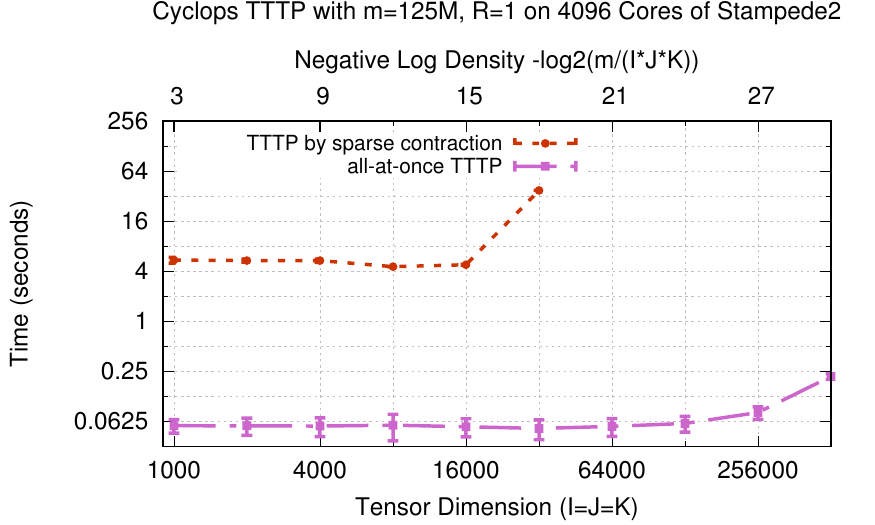}
\label{fig:tttpr1}
}\end{subfigure}
\begin{subfigure}[TTTP with $R=60$]
{\includegraphics[width=0.45\textwidth]{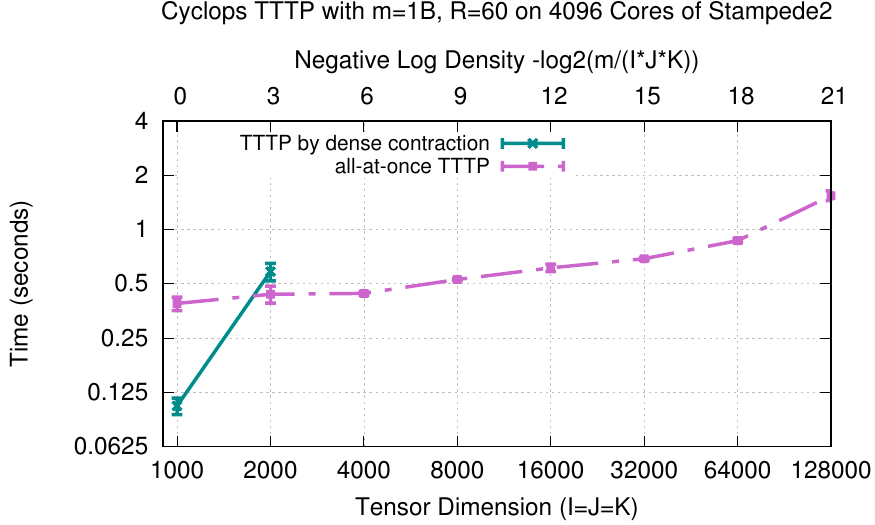}
\label{fig:tttpr60}
}\end{subfigure}
\caption{Execution time of the described TTTP kernel (all-at-once TTTP) and implementations based on pairwise tensor contraction, with $R=1$ and $R=60$ tensor products.}
}
\end{figure*}

\begin{figure}[t]
{
\centering

\includegraphics[width=0.45\textwidth]{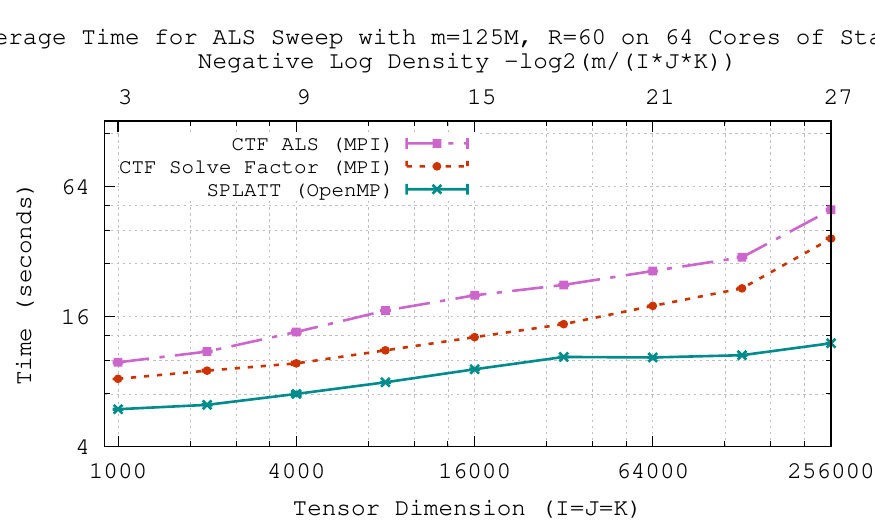}
\caption{
Comparison of performance of ALS approaches for a single ALS sweep for the random processes using 64 MPI processes for CTF and 64 OPENMP threads with 1 MPI process on a single KNL node of Stampede2.}
\label{fig:compare_als_scaling}
}
\end{figure}

We provide performance results for a range of kernels and for tensor completion algorithms overall\footnote{The tensor completion codes are available via \url{https://github.com/cyclops-community/Tensor_completion}.}.
All benchmarks and application code are written purely in Python using Cyclops without any explicit distributed data management/communication.
We study the scalability of redistribution routines within Cyclops for sparse and dense tensors by benchmarking tensor transposition and reshaping routines.
We then consider performance of the new hypersparse contraction and TTTP kernels by benchmarking TTM, MTTKRP, TTTP, and Solve Factor.
Finally, we provide a comparative study of the performance of all the algorithms introduced in Section~\ref{sec:cmpl} for tensor completion on a model low-rank dataset and on a realistic large tensor (Netflix dataset~\cite{bennett2007netflix}) with two different loss functions. 

\subsection{Benchmarking Configuration}

All results are collected on the Stampede2 supercomputer at Texas Advanced Computing Center (TACC) via XSEDE.
Stampede2 consists of 4200 Intel Knights Landing (KNL) compute nodes (each capable of a performance rate over 3 Teraflops/s) connected by an Intel Omni-Path (OPA) network with a fat-tree topology (achieving an injection bandwidth of 12.5 GB/sec).
We use Cyclops v1.5.5 built with Intel ICC compiler v18.0.2 with MKL and ScaLAPACK, Intel MPI, HPTT v1.0.5, and -O1 level of optimization.
We benchmark the MTTKRP in SPLATT v1.1.1 and use the `sc16` branch to benchmark tensor completion~\cite{Smith:2016:EOA:3014904.3014946}, using distributed MPI variants of both.
% (higher levels of optimization did not uniformly improve performance and lead in some cases lead to errors).
All experiments use 64 MPI processes per node, with 1 thread per process.
% (we do not leverage the OpenMP capabilities of Cyclops).
For all benchmarks except tensor completion, we quantify noise by displaying estimated 95\% confidence intervals.
These are centered at the arithmetic mean and have a width of four standard deviations in the observed data (first/warm-up trial ignored).

\subsection{Redistribution Performance}

Figure~\ref{fig:transp} and Figure~\ref{fig:reshape} consider the weak scalability of tensor transposition and reshaping.
These are commonly used as multidimensional array operations in NumPy Python code, so their performance is important for a range of applications.
Redistributions are substantially more costly in a distributed environment and are often the main bottleneck in Cyclops tensor contractions due to the necessity of communicating data between processes to a new processor grid mapping.
The number of nonzero elements is kept fixed across variants, but increased in proportion to the number of nodes used.
Overall, we observe good scalability in end-to-end bandwidth (computed as the number of bytes necessary to store the tensor divided by execution time) of the two operations.
The reshape performance for dense tensors can be improved, as it converts to sparse format, leveraging preservation of global ordering.
The performance is generally independent of tensor order or of the particular type of transpose/reshape.

\begin{figure*}[t]
\centering
\begin{subfigure}[Tensor Completion for Function Tensor Model Problem]
{\includegraphics[width=0.45\textwidth]{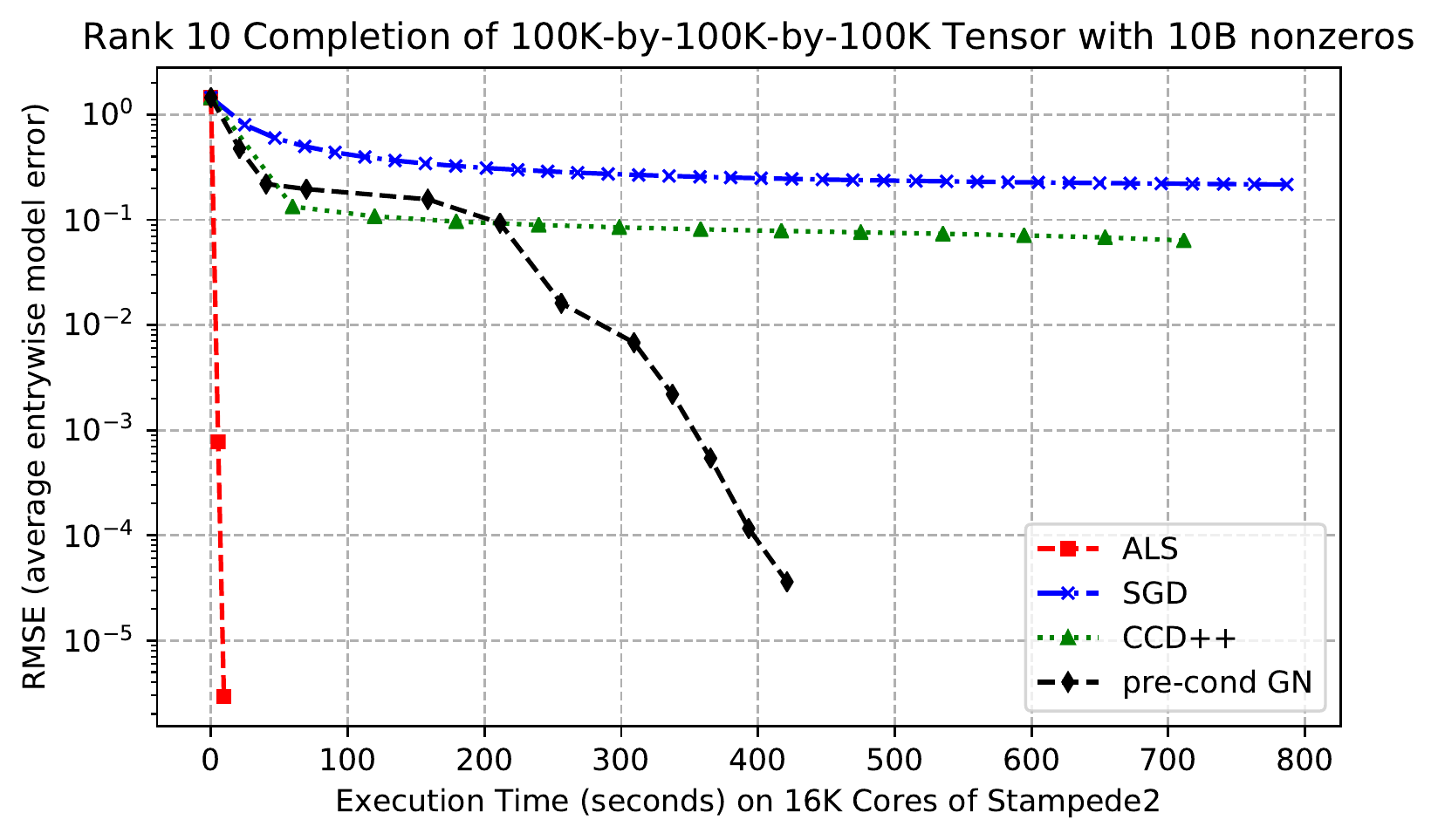}
\label{fig:cmp_func}
}\end{subfigure}
\begin{subfigure}[Tensor completion for Netflix Dataset]
{\includegraphics[width=0.45\textwidth]{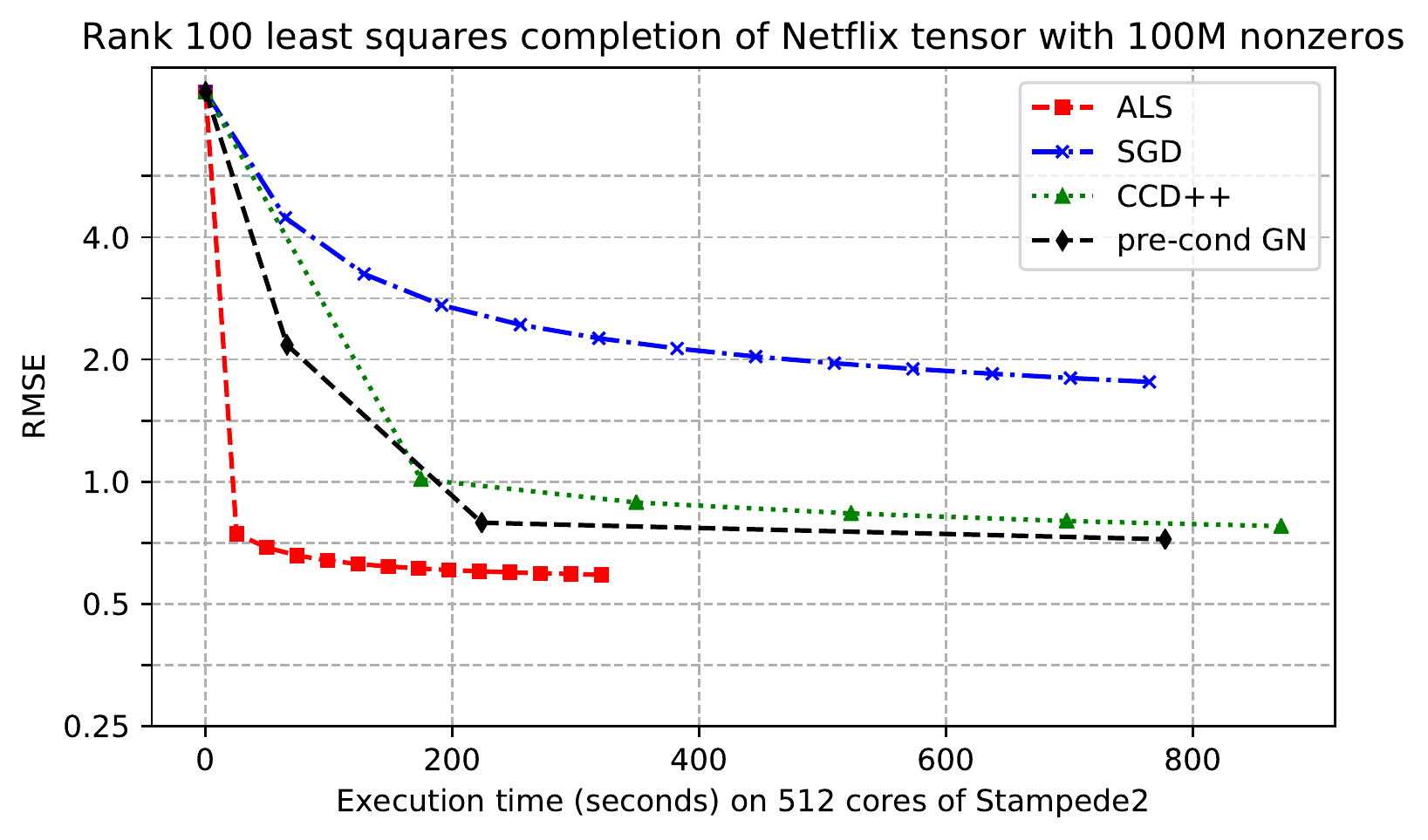}
\label{fig:cmp_netflix}
}\end{subfigure}
\begin{subfigure}[Tensor completion for tensor constructed from positive random matrices]
{\includegraphics[width=0.4\textwidth]{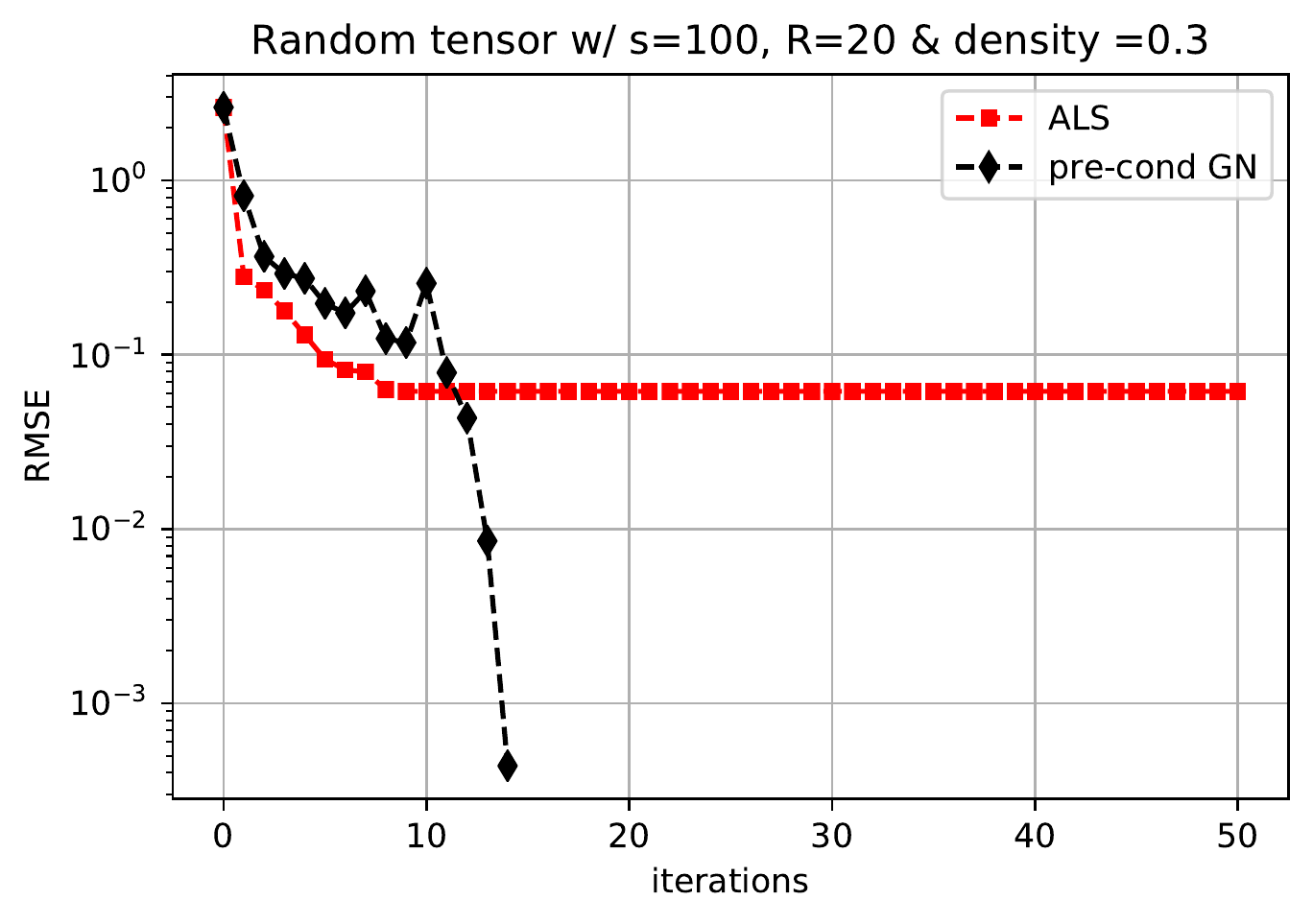}
\label{fig:cmp_random}
}\end{subfigure}
\caption{Performance results of tensor completion methods.}

\end{figure*}

\subsection{Hypersparse Representation Performance}
Figure~\ref{fig:ttm} compares variants of Cyclops tensor times matrix (TTM) kernels using 64 nodes of Stampede2 for various density of nonzeros (for a fixed nonzero count).
The performance of each variant is plotted for problem sizes for which it does not run out of memory.
We observe that the dense variant performs relatively well, but quickly runs out of memory.
Using a sparse tensor representation and a dense output representation achieves the best performance, but as the number of nonzeros grows, the output becomes sparse and representing it in a dense format incurs an unmanageable memory footprint.
Finally, the hypersparse variant, which leverages a sparse output tensor, incurs significant overhead with respect to using a dense output, but is able to scale to substantially more sparse tensors.
Overall, we conclude that the hypersparse implementation achieves the desired memory scaling, but at a significant constant factor overhead, due to the need for more sparse format conversions, indirect accesses, and sparse reduction.
%The performance could likely be improved by further optimizations, e.g., by leveraging the fact that the nonzero rows in the resulting local matrices are dense.
%Further tuning of Cyclops performance models which decide on processor grid mappings would likely lead to smoother performance trends.

Figure~\ref{fig:mttkrp} demonstrates the performance of MTTKRP using Cyclops, comparing also to the highly-optimized SPLATT implementation~\cite{smith2015tensor,smith2015splatt} (by profiling the MTTKRP within its parallel CP decomposition). In the MTTKRP kernel, a third order tensor is contracted with matrices along two modes, e.g., $\sum_{i,k} t_{ijk}u_{ir}w_{kr}$.
The given performance results are the average over the three choices of uncontracted modes.
There are two choices for performing this operation via pairwise tensor contractions, either to first contract $\CC{T}$ and $\Mat{U}$ or to first contract $\Mat{U}$ and $\Mat{W}$.
The latter can be faster if $\CC{T}$ is relatively dense, but is slower if $\CC{T}$ is sufficiently sparse.
%Cyclops makes dynamic decisions between these kernel options based on input sparsity and estimated output sparsity.
When the intermediate output tensor is sufficiently sparse and contracting with $\CC{T}$ first is estimated to take less time, Cyclops automatically leverages the hypersparse representation.
Use thereof permits scalability to much sparser tensors.
However, we observe that all-at-once computation of MTTKRP is much faster than pairwise tensor contraction.

SPLATT outperforms the Cyclops all-at-once implementation as the latter requires redistribution of factor matrices and does not use the CSF format.
However, generally Cyclops is within a factor of four or less in performance with respect to SPLATT. Further, the approach used in Cyclops permits easier combination with other tensor operations, since the input distribution of the factor matrices is not specialized for the kernel.

\subsection{TTTP Performance}
Figure~\ref{fig:tttpr1} and Figure~\ref{fig:tttpr60} compare the performance of the new TTTP kernel to alternatives based on pairwise tensor contraction, including with the use of hypersparsity.
However, even with hypersparsity, the intermediates which must be formed in any pairwise contraction tree increase the memory usage, whenever $R>1$.
We observe that the TTTP kernel is always significantly faster and can scale to extremely low density.
By comparison, pairwise tensor contraction approaches are slower even when $R=1$ and are less memory scalable.
Overall, the benefit of performing TTTP all-at-once as opposed to via pairwise contractions is clearly evident.

\subsection{Solve Factor and Alternating Least Squares Performance}

We compare our implementation of alternating minimization for least squares loss (ALS) using the Solve Factor and MTTKRP kernels introduced above to the state of the art implementation of ALS in SPLATT~\cite{Smith:2016:EOA:3014904.3014946}.
The SPLATT approach forms the left and right hand sides in a single pass over the tensor nonzero entries, thereby reusing Hadamard products of the rows for each nonzero entry. In contrast, our implementation does two passes over the tensor nonzeros.
In Figure~\ref{fig:compare_als_scaling}, we compare the performance of one ALS iteration on a tensor with fixed number of observed entries while increasing the the dimensions of the tensor dataset. SPLATT outperforms our  ALS implementation by a speed of about $2 \times$ for most of the cases. The speed up becomes $4.1 \times$ for the largest dimension because of communication among the cores in the MPI implementation, which is not needed in SPLATT's threaded implementation. Moreover, we perform a redistribution of factor matrices for each kernel call, which means a significant overhead. While somewhat slower than SPLATT, our implementation has lower memory footprint. SPLATT is unable to perform completion with MPI parallelization on one node for larger dimensions due to the memory bottleneck of forming left hand sides described in Section~\ref{sec:krnl:solvefac}. Our implementation alleviates this memory overhead by using batched computation of the rows of the required factor matrices. SPLATT stores multiple compressed sparse fibre (CSF) representations of the tensor, which eliminates the need for sorting tensor nonzeros on the fly. CSF is faster as compared to Cyclops COO-like format which requires extra computation to determine the indices for each nonzero. However, the Cyclops COO-like format is more memory efficient than SPLATT, as it uses 1 and not 3 copies of the input tensor. Further, the replicated CSF approach would entail additional overheads for generalized loss functions as the input to the kernel changes for other loss functions at each sub-iteration, necessitating construction of $3$ copies of the data at each sub-iteration. Our Solve Factor kernel is only about $1.7 \times$ slower than SPLATT for most of the cases,
suggesting the overheads of using general kernels is not too high.  

\subsection{Tensor Completion With Least Squares Loss}

Figure~\ref{fig:cmp_func} studies the performance of tensor completion algorithms with Cyclops on a model problem constructed from a sampled function as described in Karlsson et al~\cite{KARLSSON2016222}. The sampled tensor has low CP rank (we pick $R=10$) and a good CP decomposition is easily found by quadratic approximation. We observe that ALS requires only a few iterations to achieve full accuracy (RMSE proportional to the regularization used, $\lambda=10^{-5}$). CCD++ is executed with a regularization of $\lambda=10^{-5}$ and SGD is executed with sampling and learning rate of $5 \cdot 10^{-3}$ and regularization of $10^{-7}$. The CCD++ and SGD approaches achieve comparable performance, requiring less time per iteration, but making progress at a slower rate overall (RMSE plotted after every 20 iterations).  Pre-conditioned Gauss-Newton method also converges to full accuracy in a few iterations (regularization used $\lambda=10^{-3}$) but is considerably slower than ALS execution time. Using 256 nodes of Stampede2, this experiment demonstrates the scalability of our Python-based tensor completion implementations, as they are executed on a problem containing 10 billion observed entries (nonzeros) with a density of $10^{-5}$.

In Figure~\ref{fig:cmp_netflix}, we consider performance for the Netflix movie rating dataset on 4 nodes of Stampede2 with a rank $100$ CP representation.This tensor is $480,189\times 17,770\times  2,182$ and contains $m=100,477,727$ nonzeros. While ALS achieves the lowest RMSE, the three methods that use second order information are relatively competitive for this tensor. ALS iterations take the least time followed by the CCD++ iterations, For CCD++, we traverse the tensor nonzeros $2R$ times for each CCD++ iteration as compared to $2$ times for each ALS iteration. Both algorithms use a regularization parameter of $\lambda = 10^{-5}$. Gauss-Newton with implicit pre-conditioned CG uses a relative tolerance of $5 \cdot 10^{-3}$ and max iterations of at most $30$ for CG and a regularization parameter $\lambda = 10^{-3}$. We use the Solve Factor kernel to implement block diagonal pre-conditioning which is essential for faster convergence and stability of CG iterations.  The algorithm starts to take more time as CG iterations start to increase due to the fact that gradient norm decreases. Unlike the function tensor model problem, SGD requires fine-tuning of parameters, diverging when the learning rate is set to be too high. We show performance with a learning and sampling  rate of $3\cdot 10^{-3}$ with $\lambda = 10^{-5}$, which resulted in cheap iterations and steady but slow convergence (RMSE plotted after every 20 iterations). The progress made by the SGD steps can likely be improved by strategies that vary the learning and sample rate, a consideration which we leave for future work. 

In Figure~\ref{fig:cmp_random}, we consider performance for pre-conditioned Gauss-Newton method and ALS for a synthetic tensor. This tensor is constructed with random matrices with entries sampled uniformly from $[0,1]$ with dimension $s=100$ and CP rank $R=20$ with $30\%$ observed entries, i.e., the tensor has $3 \cdot10^{5}$ observed entries and is relative dense. We observe that for this type of problem, pre-conditioned Gauss-Newton converges to the solution in a few iterations whereas ALS seems to make very little progress after $10$ iterations. This corroborates the claim in the previous work~\cite{liu2020tensor} that ALS does not perform well for relatively dense tensors and methods like Gauss-Newton may be preferable when an exact solution exists.
%
%We show the performance of ALS, CCD++, and SGD with and without the all-at-once MTTKRP kernel.
%The all-at-once kernel provides a clear benefit for all three methods.
%
%When used without MTTKRP, the SGD regularization is decreased by $.99$ at each step.
%
%SGD is executed with a sample and learning rates of $10^{-3}$.
%

%

%For CCD++, we observe a 1.40X and 1.84X speed-up in time per iteration from using the TTTP-based implementation for the function tensor model problem and Netflix dataset, respectively.
%The sparse summations performed in the TTTP-based implementation are observed to be a more significant bottleneck than the TTTP routine.
%Overall, the TTTP routine is more robust in performance than the sparse and hypersparse tensor contraction kernels, and we conclude that it can outperform hypersparse tensor contractions for a sparse rank-1 MTTKRP.

%Overall, the results demonstrate the capability of the Cyclops-based parallel tensor completion implementations to leverage over ten thousand processes concurrently, handle 10 billion nonzeros with a sparsity of 1 in 100K, as well as to handle a non-equidimensional real-world tensor dataset (Netflix).
%While hypersparse kernels enable execution of tensor completion with massively-sparse tensors, all-at-once MTTKRP is much faster.
%and MTTKRP can likely be improved, our approach realizes effective asymptotic cost and memory scaling, enabling the use of distributed memory for massively large datasets with little implementation effort.

\subsection{Tensor Completion With Poisson Loss}

To demonstrate our algorithmic and software framework for generalized CP completion, we implement the above described algorithms for tensor completion with Poisson loss with the logarithm link function (log-link) for the Netflix tensor. Poisson loss for decomposing tensors with entries in the set of natural numbers has several qualitative advantages that have been explored in the previous literature ~\cite{hansen2015newton,chi2012tensors}. We explore the quantitative performance and scalability of various algorithms in a distributed setting.

Poisson loss with log-link was introduced in~\cite{hong2020generalized} for tensors with entries in the set of natural numbers. The advantage of using log-link is that it relaxes the nonnegativity constraints required with the identity-link and hence, we can use our framework to implement all the algorithms without having to account for any constraints. The loss function minimized here is described by setting the elementwise function $\phi$ introduced in equation~\ref{eq:obj} to
\begin{align*}
\phi(t_{ijk},\langle \Vec{u}_i, \Vec{v}_j, \Vec{w}_k\rangle) =  \exp{(\langle \Vec{u}_i, \Vec{v}_j, \Vec{w}_k\rangle)} - t_{ijk}\langle \Vec{u}_i, \Vec{v}_j, \Vec{w}_k\rangle.
\end{align*}

\begin{figure*}[t]
\centering
\begin{subfigure}[Comparison of performance of different Poisson tensor completion algorithms]
{\includegraphics[width=0.45\textwidth]{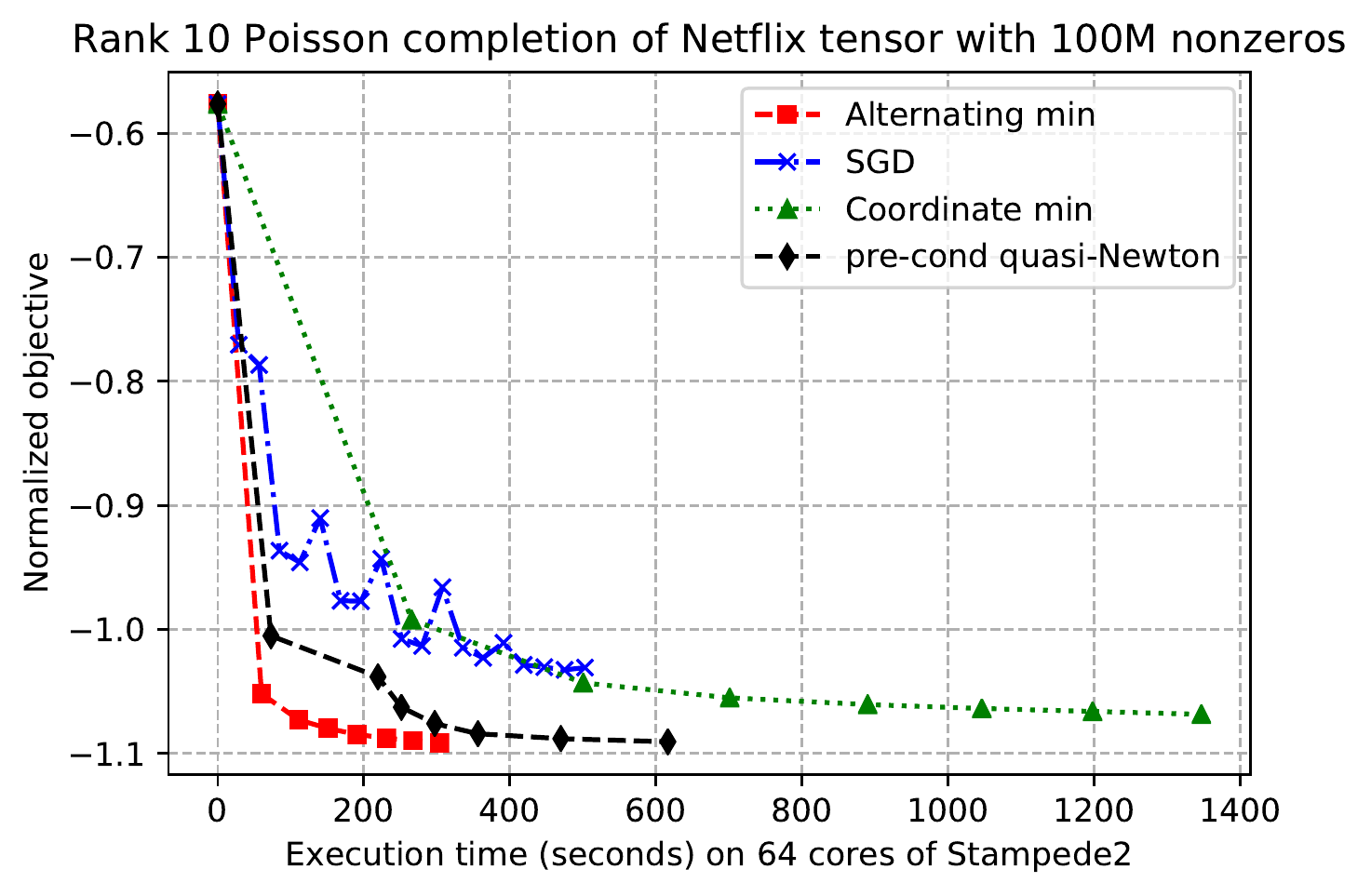}
\label{fig:poisson_netflix}
}\end{subfigure}
\begin{subfigure}[Comparison of performance of alternating minimization on different loss functions]
{\includegraphics[width=0.45\textwidth]{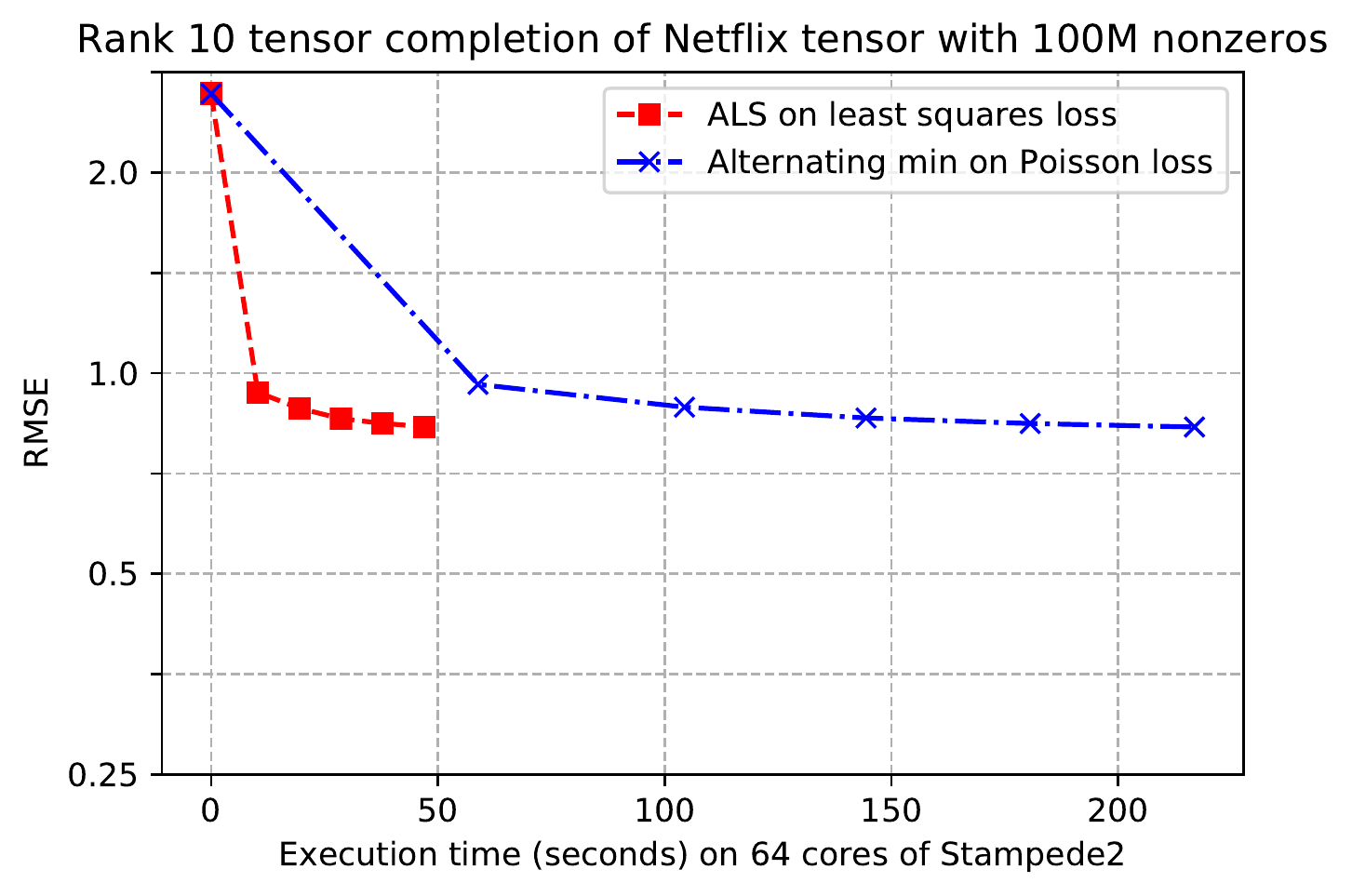}
\label{fig:poisson_netflix_time}
}\end{subfigure}
\caption{Tensor completion with Poisson and least squares loss function on Netflix tensor on 64 cores of Stampede2}
\end{figure*}

With the elementwise function defined as above, we use values from Table~\ref{tab:derivs} to implement all the completion algorithms described in Section~\ref{sec:cmpl} with this loss for the Netflix tensor with rank $R = 10$. We plot the normalised loss, i.e., $\frac{1}{|\Omega|}\sum_{i,j,k} \phi(t_{ijk},\langle \Vec{u}_i, \Vec{v}_j, \Vec{w}_k\rangle)$ versus time for each algorithm in Figure~\ref{fig:poisson_netflix}. Each point in the Figure~\ref{fig:poisson_netflix} represents an iteration, except for SGD, for which each point is plotted after every 20 iterations. Both the alternating minimization and coordinate minimization inner iterations are performed until a relative step tolerance of $10^{-3}$ or a maximum count of $5$ is reached. Pre-conditioned quasi-Newton has a relative tolerance of $5 \cdot 10^{-3}$ or a maximum iteration count of $R$ for CG iterations for each system solve. Also, the regularization parameter plays a pivotal role for this objective function as algorithms diverge easily due to the exponentiation. Compared to least-squares loss, we employ higher values of regularization for all the algorithms to ensure that they do not diverge. We use a value of $\lambda= 1$ for coordinate minimization, and $\lambda = 0.1$ for alternating minimization, quasi-Newton, and SGD. 

We observe that alternating minimization is the fastest algorithm to reach the least value of the objective followed by pre-conditioned quasi-Newton method and then coordinate minimization. While implementation of all these algorithms can be fine-tuned to further run faster for the particular loss functions, we observe that SGD implementation is outperformed by other algorithms indicating that the benefit of using second order information. Our formulation of the quasi-Newton with implicit pre-conditioned CG method not only makes the implementation feasible via tensor algebra kernels, but is also competitive with other algorithms in  practical scenarios.

For a Poisson loss objective, we plot the Frobenius norm of the subtraction of input tensor and exponentiated reconstructed tensor at each iteration and compare it with the ALS iterations. We observe that these values are equal up to 2 digits suggesting that the Poisson loss also minimizes the least squares loss, however, vice versa is not true as there may be negative values making the Poisson objective infeasible to calculate. Note that the Poisson loss completion comes at the cost of performing inner iterations for solving each factor, which results in longer running time, as observed in Figure~\ref{fig:poisson_netflix_time}.

\section{Related Work}
\label{sec:rel}

We review related work on parallel tensor abstractions and on previous parallel implementations of tensor completion. We also review work on sparse tensor kernels for tensor decompositions.

\subsection{Parallel Tensor Completion}

The tensor completion algorithms presented in this paper have commonly-used analogous in matrix completion (ALS~\cite{jain2013low}, SGD~\cite{keshavan2010matrix}, CCD~\cite{yu2012scalable}).
These approaches, especially SGD, have been optimized extensively for the matrix case, which may be viewed as a simple two-layer neural network.
In shared memory, SGD is widely used, as it can be made efficient by asynchronous execution~\cite{recht2011hogwild}.
ALS, CCD, and SGD for matrix completion have all been target of efficient distributed-memory implementations~\cite{teflioudi2012distributed,yu2012scalable,hastie2015matrix,gemulla2011large}.

Tensor completion via the CP tensor representation~\cite{gemulla2011large} has been a target of recent distributed-memory implementation efforts.
Karlsson et al.~\cite{KARLSSON2016222} implement ALS and CCD by replicating the factor matrices on each process and distributing observed entries.
While efficient, this approach is not scalable to very large factor matrices.
Smith et al.~\cite{Smith:2016:EOA:3014904.3014946} improve upon this method by distributing both the factor matrix and tensor in coherent formats, similar to our parallel method for TTTP when it is done with a single parallel step.
Our work is the first to implement distributed tensor completion using high-level tensor operations for general tensor contractions.
We reproduce previous work~\cite{KARLSSON2016222,Smith:2016:EOA:3014904.3014946} in the observation that ALS is generally most efficient for distributed tensor completion.

\subsection{Sparse Tensor Kernels}

Parallel sparse matrix multiplication algorithms comprise an active area of research~\cite{Solomonik:2017:SBC:3126908.3126971,koanantakool2016communication,Ballard:2013:COP:2486159.2486196,ballard2015brief,DBLP:journals/corr/abs-1109-3739,gustavson1978two,10.1007/978-3-662-44777-2_62}.
Multiplication of hypersparse matrices has seen considerably less study~\cite{buluc2008representation}.
An optimized doubly compressed CSR/CSC layout eqivalent to the CCSR layout used in this paper is the standard sequential approach to hypersparse matrix--matrix products~\cite{buluc2008representation}.

Effective sparse tensor layouts have been designed for TTM and MTTKRP operations in shared memory and distributed memory.
The compressed sparse fiber (CSF) layout serves as an extension of hypersparse matrix representations and achieves efficient storage and TTM operations~\cite{smith2015splatt,smith2015tensor}.
The hierarchical coordinate (HiCOO) layout is designed to further improve efficiency for TTM and MTTKRP~\cite{li2018hicoo}.
The tensor algebra compiler (TACO) supports hierarchical layouts with compressed or uncompressed modes~\cite{kjolstad2017tensor} as well as other optimized sparse formats~\cite{chou2018formats}.
These layouts can be interchanged and may improve upon the CCSR layout used in our work.
However, our design is the first to enable arbitrary tensor contractions to be reduced to a storage-efficient layout, and to support distributed-memory tensor operations with hypersparse representations.

TTM and MTTKRP are standard benchmark tensor kernels~\cite{doi:10.1137/07070111X,li2019pasta}.
MTTKRP has been the target of optimization for distributed-memory architectures with both MPI~\cite{smith2015splatt,kaya2015scalable} and MapReduce~\cite{park2016bigtensor,blanco2018cstf}.
While TTM is a special case of a tensor contraction, MTTKRP involves contraction of multiple tensors and consequently presents potential for further performance optimization over pairwise contraction by all-at-once contraction~\cite{hayashi2018shared}.
The TTTP operation introduced in this paper differs significantly from MTTKRP and can be specially optimized via all-at-once contraction.
%a potential standard tensor benchmark.

\subsection{Tensor Frameworks}

Tensors and multidimensional arrays are a prevalent programming abstraction that encapsulates data parallelism.
Many tensor libraries are designed for methods in quantum chemistry.
%, which motivated the development of Cyclops, has also been the driving application for a number of other tensor efforts.
The Tensor Contraction Engine (TCE)~\cite{doi:10.1021/jp034596z} provides factorization of multi-tensor expressions into pairwise contractions.
TCE generates parallel tensor contraction code based on a partitioned global address-space (PGAS)~\cite{yelick2007productivity} language, Global Arrays~\cite{springerlink_ga1}.
Global Arrays and other PGAS languages such as UPC~\cite{el2005upc} provide multidimensional array abstractions that enable tensor programming, but generally do not support high-level tensor algebra operations.
The Libtensor library~\cite{libtensor} provides efficient shared-memory tensor contractions, targeted at quantum chemistry applications.
Libtensor and other libraries~\cite{mutlu2019toward} support block-sparse tensors.
The TiledArray~\cite{peng2016massively,calvin2015scalable} library provides distributed-memory support for block-sparse tensor contractions.
Outside of Cyclops, to the best of our knowledge, tensor contractions with arbitrary elementwise sparsity are only supported for single-node execution~\cite{Kats_sp_tensor2013}.
The above efforts all leverage an Einstein notation syntax for contractions and aim at efficient execution of tensor contractions arising in quantum chemistry.

The Tensor Algebra Compiler (TACO)~\cite{kjolstad2017tensor} provides support for sequential sparse tensor contractions and more general multi-tensor expressions.
In recent work, TACO has been improved to factorize longer tensor algebra expressions and their subcomponents into subsequences~\cite{kjolstad2019workspaces}, the former being a user-guided version of the automated factorization in Cyclops.
Tensor libraries have also been designed for machine learning workloads, e.g., TensorFlow by Google~\cite{199317} and Tensor Comprehensions by Facebook~\cite{vasilache2018tensor}.
Both focus on task-level parallelism and GPU acceleration as opposed to distributed-memory data parallelism.

%\begin{itemize}
%    \item Other tensor/multidimensional-array programming libraries, parallel, sparse, Einstein summation
%    \item Previous studies of tensor completion methods and parallel performance \\
%    % issue 3 Paragraph summarizing Kressner et al wor
%    The parallel formulation of tensor completion algorithms on ALS and CCD++ in the CP format proposed by ~\cite{KARLSSON2016222} distribute variables on each processor and update each subset of variables independently. 
%    The rows of current factor matrix are updated simultaneously and summed together by inter-process communication in the massage-passing paradigm.
%    Both ALS and CCD++ algorithms are weakly scalable. Analysis on synthetic and real-world data indicate that ALS is more sensitive to tensor CP rank and converge faster compared to CCD++.
%
%    % issue 4 Paragraph summarizing Shaden's paper and other CCD-related work
%    % add more: how ALS,SGD,CCD++ parallelize
%    Shaden et al. [Ref] study three optimization algorithms for tensor completion: ALS, SGD, and CCD++, which are compared against each other by convergence rates. 
%    SGD is most competitive in a serial environment, ALS is advantageous on shared-memory systems, while both ALS and CCD++ are competitive on distributed systems.
%\end{itemize}

\section{Conclusion and Future Work}
\label{sec:cnc}

We present new advances in parallel sparse tensor computations infrastructure and methodology, driven by its application to tensor completion.
Specifically, we propose a new tensor algebra routine, TTTP, which consists of tensor contractions that may be significantly accelerated by an all-at-once contraction algorithm.
Further, we provide the first distributed general sparse tensor contraction infrastructure that can leverage hypersparse matrix representations, achieving scalability to massively sparse tensors.

For tensor completion, we propose a novel Newton-method-based algorithmic framework for generalized tensor completion. In this framework, we introduce alternating minimization, coordinate minimization and quasi-Newton algorithms which encompass the ALS, CCD++ and Gauss-Newton algorithm for least squares loss and generalize easily for other objective functions. Our results demonstrate that these algorithms are more accurate than the SGD algorithm for generalized completion. By providing a high-level Python interface to the tensor algebra operations, we are able to develop very concise, but massively-parallel implementations of these algorithms for generalized tensor completion via CP decomposition. Moreover, we show that our distributed memory implementation of alternating minimization for least squares loss is competitive with the state of the art distributed implementation of ALS.
Our experimental results demonstrate that hypersparsity, all-at-once kernels for MTTKRP, and the new TTTP algorithm enable generalized tensor completion algorithms to be executed on much sparser tensors than possible with previously available libraries.

For the generalized objective functions, some of the link functions use nonnegativity constraints ~\cite{hong2020generalized}, which are not incorporated in our current framework. While all the link functions can be modified to remove these constraints, the interpretation of the factors might change. These constraints can be incorporated with use of projected Newton's algorithm~\cite{bertsekas1982projected} or using a barrier formulation. All the kernels introduced in Section~\ref{sec:krnl} can be optimized further by using specialised tensor formats like CSF coupled with an optimal threaded implementation for best performance. However, it is non-trivial to construct these formats optimally for each sub-iteration for a generalized loss functions.

%\section{Acknowledgements}
%\label{sec:ack}
%\input{ack}

\begin{acks}
This work used the Extreme Science and Engineering Discovery Environment (XSEDE), which is supported by National Science Foundation grant number ACI-1548562. Via XSEDE, the authors made use of the TACC Stampede2 supercomputer.
The research was supported by the US NSF OAC via award No.\ 1942995.
\end{acks}

\bibliographystyle{ACM-Reference-Format}
\bibliography{paper}

\appendix 

\section{Appendix}
\label{sec:additional}

\subsection{Generalized CP decomposition}

All the algorithms for generalized CP completion introduced in Section~\ref{sec:cmpl} are based on elementwise derivatives of the generalized objective function~\cite{hong2020generalized} with respect to each variable in the factor matrix. In this section, we use tensor calculus to derive the necessary expressions for an $N^{\text{th}}$ order input tensor $\CC{X} \in \mathbb{R}^{I_1 \times \dots \times I_N}$. We assume that we have an index set $\Omega \subset \{1,\dots,I_1\} \times \dots \times \{1,\dots I_N\}$, which represents the set of observed entries of the input tensor. If $\Omega$ consists of all the elements then the objective function would correspond to a decomposition problem. The objective function is 
\[f(\mathbf{A}^{(1)} \ldots \mathbf{A}^{(N)}) = \sum_{i_1,\ldots,i_N \in \Omega}\phi(x_{i_1 \ldots i_N},m_{i_1 \ldots i_N}), \quad \quad \text{where} \quad m_{i_1\ldots i_N} = \sum_{r=1}^R\prod_{n=1}^Na^{(n)}_{i_nr}. \]
The elementwise expression for the gradient of $f(\mathbf{A}^{(1)} \ldots \mathbf{A}^{(N)} ) $ with respect to $d^{\text{th}}$ factor matrix is
\[ \frac{\partial f}{\partial a^{(d)}_{kr}} = \sum_{i_1,\ldots,i_N \in \Omega}  \frac{\partial \phi(x_{i_1\ldots i_N},m_{i_1\ldots i_N})}{\partial m_{i_1\ldots i_N} } \delta_{i_dk}\prod_{n=1,n \neq d}^Na^{(n)}_{i_nr}.\]
Computing the gradient corresponds to an MTTKRP operation with the derivative tensor $\phi'_{i_1 \dots i_N}$ (Equation~\ref{eq:deriv_tnsr}). We can differentiate the above expression for gradient further to arrive at an elementwise form of the Hessian matrix. This form is useful for writing algorithms that use second order information as these use a part of the Hessian matrix and/or use the implicit form of the Hessian. The derivative of the gradient with respect to $p^{\text{th}}$ factor matrix can be calculated by applying chain rule inductively 
\begin{align*}
h^{(d,p)}_{krlz}=\frac{\partial f^2}{\partial a^{(d)}_{kr} \partial a^{(p)}_{lz}}&=
\sum_{i_1,\ldots,i_N \in \Omega}\phi''_{i_1\ldots i_N}\delta_{i_pl}\bigg(\prod_{n=1, n\neq p}^Na^{(n)}_{i_nz}\bigg)\delta_{i_dk}\bigg(\prod_{n=1, n\neq d}^Na^{(n)}_{i_nr}\bigg)\\ 
&+(1- \delta_{dp})\sum_{i_1,\ldots,i_N \in \Omega}\phi'_{i_1\ldots i_N}\delta_{i_dk}\bigg(\prod_{n=1, n\neq d,p}^Na^{(n)}_{i_nr}\bigg)\delta_{i_pl}\delta_{rz},
\end{align*}
where $\delta_{ij}$ is the Kronecker-Delta function. Newton or quasi-Newton method require solution to  linear systems  involving the Hessian at each iteration. A Krylov subspace method can be used to solve these system of equations by making use of the implicit form of the Hessian. Given current factor matrix updates $\Mat{W}^{(1)} ,\ldots, \Mat{W}^{(N)}$, the matrix-vector product with the Hessian can be computed by the following tensor contractions,
\[w^{(d)(new)}_{kr} =\sum_{p}\sum_{l,z}h_{krlz}^{(d,p)}w^{(p)}_{lz}, \quad d \in \{1, \ldots, N\}, p \in \{1,\ldots, N\}, \]
where $\Mat{W}^{(d)(\text{new})}$ is the update matrix corresponding to the $d^{\text{th}}$ factor matrix. These contractions reduce to simpler contractions as mentioned in Section~\ref{sec:cmpl:GN}. The above form of Hessian can be used to derive all the methods described in Section~\ref{sec:cmpl}.

Alternating minimization described in Section~\ref{sec:cmpl:ALS} is equivalent to a block non-linear Gauss-Siedel method~\cite{grippo2000convergence} to minimize the above objective function. Alternating minimization subiteration uses a diagonal block of the above described Hessian for optimizing a factor matrix given by 
\[h^{(d,d)}_{krlz}=
\sum_{i_1,\ldots,i_N \in \Omega}\bigg(\prod_{n=1, n\neq d}^Na^{(n)}_{i_nz}\bigg)\delta_{i_dk}\phi''_{i_1\ldots i_N}\delta_{i_dl}\bigg(\prod_{n=1, n\neq d}^Na^{(n)}_{i_nr}\bigg).\]
Coordinate minimization subiteration described in Section~\ref{sec:cmpl:CCD} is equivalent to a non-linear Gauss-Seidel method, as in each subiteration, the method minimizes only one variable (in parallel) at a time. It uses the diagonal of the diagonal block of the above described Hessian given by
\[h^{(d,d)}_{kr}=
\sum_{i_1,\ldots,i_N \in \Omega}\bigg(\prod_{n=1, n\neq d}^Na^{(n)}_{i_nr}\bigg)^2\delta_{i_dk}\phi''_{i_1\ldots i_N}.\]

\end{document}